# Picture-word interference in language production studies: Exploring the roles of attention and processing times

Audrey Bürki & Sylvain Madec

University of Potsdam, Karl-Liebknecht-Straße 24-25, 14476 Potsdam, Germany

buerki@uni-potsdam.de, symadec@gmail.com

## Abstract

The picture-word interference paradigm (participants name target pictures while ignoring distractor words) is often used to model the planning processes involved in word production. The participants' naming times are delayed in the presence of a distractor (general interference). The size of this effect depends on the relationship between the target and distractor words. Distractors of the same semantic category create more interference (semantic interference), distractors overlapping in phonology create less interference (phonological facilitation). The present study examines the relationships between these experimental effects, processing times, and attention in order to better understand the cognitive processes underlying participants' behavior in this paradigm. Participants named pictures with a superimposed line of Xs, semantically related distractors, phonologically related distractors, or unrelated distractors. General interference, semantic interference, and phonological facilitation effects are replicated. Distributional analyses reveal that general and semantic interference effects increase with naming times, while phonological facilitation decreases. The phonological facilitation and semantic interference effects are found to depend on the synchronicity in processing times between the planning of the picture's name and the processing of the distractor word. Finally, EEG power in the alpha band before stimulus onset varied with the position of the trial in the experiment and with repetition but did not predict the size of interference/facilitation effects. Taken together, these results suggest that experimental effects in the picture-word interference paradigm depend on processing times to both the target word and distractor word and that distributional patterns could party reflect this dependency.

**Picture-word interference in language production studies: Exploring the roles of attention and processing times**

Audrey Bürki & Sylvain Madec

University of Potsdam, Karl-Liebknecht-Straße 24-25, 14476 Potsdam, Germany

buerki@uni-potsdam.de

**Introduction**

The picture-word interference paradigm is a very popular paradigm in language production research. In this paradigm, participants are asked to name pictures while ignoring other words, i.e., distractors, presented visually or orally. This paradigm is regularly used to test key aspects of the cognitive architecture of the language production system, including the nature of lexical access (e.g., Abdel Rahman & Aristei, 2010; Damian & Bowers, 2003; Roelofs, 1992; Starreveld et al., 2013), the involvement of inhibition in language production (e.g., Shao et al., 2015), the scope of advanced planning (e.g., Meyer, 1996; Michel Lange & Laganaro, 2014), or the time course of processes (e.g., Schriefers et al., 1990). The paradigm has been used with a variety of populations, including older individuals (Taylor & Burke, 2002), bilinguals (Roelofs et al., 2015; Sá-Leite et al., 2020), or adults and children with language impairments (Hashimoto & Thompson, 2010; Seiger-Gardner & Schwartz, 2008). Despite the popularity of the paradigm, the cognitive mechanisms underlying the participants behavior in this task are not fully understood, and oft debated (Bürki et al., 2020; Dell'Acqua et al., 2007; Starreveld & La Heij, 2017; van Maanen et al., 2008).

Several experimental effects have been described. Naming a word in the context of another word is costly: response (or naming) latencies to the target picture are slower with a distractor than with a line of Xs, or a sequence of non-meaningful symbols (i.e., general interference effect, see for instance Glaser & Düngelhoff, 1984; La Heij & Vermeij, 1987; Lupker, 1982; Starreveld & La Heij, 1995). The nature of the relationship between distractor and target words further matters. When the distractor is of the same semantic category as the target word (e.g., *apple-banana*), naming latencies are slower than when the two words are unrelated (e.g., *apple-chair,* semantic interference effect, e.g., Lupker, 1979; Rosinski, 1977 for the first demonstrations of this effect, see Bürki et al., 2020 for a recent meta-analysis). When the two words overlap in orthography or phonology, naming latencies are shorter than with an unrelated distractor (phonological facilitation effect, e.g., Bi, Xu, & Caramazza, 2009; Damian & Martin, 1999; Posnansky & Rayner, 1977; Rayner & Posnansky, 1978; Starreveld & La Heij, 1996; de Zubicaray et al. 2002).

In the present study, our aim is to better understand the cognitive mechanisms underlying participants' behavior in the picture-word interference paradigm by examining the links between experimental effects, processing speed, and the participants' attentional state prior to stimulus onset. We study the distributional properties of experimental effects, seek experimental evidence supporting the assumption that experimental effects depend on the synchronization between the processing of the target and that of the distractor word, and examine the hypothesis that these experimental effects are modulated by the participants' attentional state.

In the remainder of the introduction we first discuss the role of processing times in picture-word interference effects. We then summarize the findings of previous studies regarding the distributional properties of such effects. We then discuss the role of attention in the Stroop paradigm before presenting the current study in some detail.

*Timing in the picture-word interference paradigm*

Experimental effects in the picture-word interference paradigm depend on the relative time course of picture and distractor *presentation*. For instance, Schriefers et al. (1990) reported semantic interference when the distractor was presented 150 ms before the picture (negative Stimulus Onset Asynchrony or SOA), and phonological facilitation (but no semantic interference) when the distractor was presented 150 ms after the picture (positive SOA, see also Damian & Martin, 1999; Glaser & Düngelhoff, 1984; Rizio et al., 2017; Taylor & Burke, 2002; de Zubicaray & McMahon, 2009).

Interactions between experimental effects and SOA are usually taken to indicate that these experimental effects depend on the relative timing of *processing* of the two stimuli. Distractors can only interfere with (or facilitate) the preparation of the target word if the relevant information is activated at the right time. If the distractor is processed too early, the corresponding activation has decayed by the time the target word is encoded, and if the distractor is processed too late, this information becomes activated too late to interfere. The observation that semantic interference occurs at earlier SOAs than phonological facilitation is argued to be in line with the assumption of many word production models, that semantic and phonological information are encoded sequentially (Schriefers et al., 1990, but see Miozzo et al., 2014).

Bürki (2017) tested the claim that phonological facilitation in the picture-word interference paradigm depends on the relative timing of *processing* of the two stimuli. In this study, the participants saw the same picture-word stimuli twice. The first time they named the pictures, the second time they read the words aloud. A measure of the difference in processing speed for the target and distractor words was computed by subtracting, for each picture-word pair, the reading time for the distractor from the naming time for the target picture. The phonological facilitation effect was found to interact with the difference measure; in line with the hypothesis that this effect only occurs for trials for which the activation of the phonological content of the target and distractor words is synchronized. In that study, the picture naming and distractor naming latencies were collected on trials in which both the picture and the distractor were presented. Moreover, the naming latencies for the same set of trials were used to compute the difference measure and as dependent variable. A more stringent test of the hypothesis that the phonological facilitation effect depends on the relative timing of processing for the target and distractor words could be achieved with a difference measure computed on independent trials, using pictures without distractors to collect naming latencies and written words without pictures to collect reading latencies. To our knowledge, no study examined whether other experimental effects, such as the semantic interference effect, interact with a measure of the relative timing of target and distractor processing (but see Piai et al., 2015 for an attempt to relate the semantic interference effect in a dual task to the participants' reading abilities).

The role of synchronization between target and distractor word processing in the picture-word interference paradigm has been discussed in the context of the distractor frequency effect, i.e., the finding that more frequent distractors create less interference than less frequent distractors. Geng et al. (2014) noted that the frequency of the distractor influences naming times in the picture-word interference paradigm but does not do so in the classical Stroop task (i.e., participants see printed words in different colors and their task is to name the color of the hue, Stroop, 1935). Geng et al. (2014) further noted that naming latencies in the Stroop task are usually shorter than in the picture-word interference task. They hypothesized that in the Stroop task, the distractor might be processed too late to interfere with the planning of the target word. To test this hypothesis, they manipulated the design of Stroop and picture-word experiments to generate either a speed up or a slowdown in overall naming times. They reported a distractor frequency effect in versions of the Stoop task where naming times were slowed down, and no effect in versions of the picture-word interference task where naming latencies were speeded up. Geng et al. (2014) take these findings to support a temporal account of the distractor frequency effect (see also Miozzo and Caramazza, 2003). Accordingly, high frequency distractors are processed -and therefore activated- earlier than low frequency distractors. By the time the target word is selected, their activation has already decayed. Note that the experiments in Gheng et al. (2014) do not provide direct empirical evidence that synchronization in processing times between target and distractor words modulate the distractor frequency effect. Only the processing times for the pictures are reported to vary across experiments. Moreover, it remains to be shown whether the appearance / disappearance of the distractor frequency effect can be directly related to picture naming speed, or is a consequence of the changes implemented in the design to modulate naming speed. Other accounts of the distractor frequency effect are discussed in the literature. According to the input account, (see Miozzo and Caramazza, 2003), the effect can be related to the duration of distractor processing alone and Starreveld et al. (2013) discuss an account where the distractor frequency effect is caused by differences in the degree of activation between frequent and infrequent distractors. Crucially, the demonstration that the distractor frequency effect depends on the relative timing of processing between target and distractor words would not necessarily mean that effects such as semantic interference or phonological facilitation also depend on this synchronization.

In sum, whereas it is often implicitly assumed that experimental effects in the picture-word interference paradigm depend on the synchronization of distractor and target word processing times, there is little direct empirical evidence supporting this claim. A fair understanding of the relationship between experimental effects and processing times in this paradigm is crucial for theoretical and methodological reasons. From a theoretical perspective, this relationship can inform the cognitive mechanisms underlying experimental effects. From a methodological perspective, understanding this relationship is crucial to determine the extent to which changes in the experimental design influence experimental effects directly or via their impact on processing times. Interference and facilitation effects in the picture-word interference paradigm are regularly used as a tool to examine other issues, such as the role of cognitive resources in language production (e.g., working memory, see for instance Klaus et al., 2017), the scope of advanced planning (e.g., Meyer, 1996), or the role of response relevance, to cite a few. To address these issues, experimental effects are compared across conditions (e.g., with different types of concomitant tasks, e.g., Klaus et al., 2017; with different types of responses, e.g., bare nouns vs. noun phrases, e.g., Klaus & Schriefers, 2018; with and without familiarization with the picture stimuli, e.g., Gauvin et al., 2018). Interactions between interference or facilitation and these conditions are used to inform the issues at stake. However, many of these manipulations have an impact on naming times (and possibly also on distractor processing times) which could, alone, explain why interference or facilitation effects vary across conditions. The demonstration that experimental effects in the picture-word interference paradigm depend on the temporal alignment of target and distractor word processing would require that the role of processing times be considered when interpreting interactions between design manipulations and interference/facilitation effects.

*Distributional analyses of (semantic) interference effects*

The relationship between experimental effects and naming times was also examined with distributional analyses. In these studies (most of them on the semantic interference effect), the naming times of each participant are ordered from fastest to slowest and then divided in quantiles. The aim is to determine whether an experimental effect is constant across quantiles or changes (increases or decreases) as naming times increase. Previous studies all report that semantic interference becomes greater when naming times increase, but they disagree as to whether the effect is found over the entire distribution of naming times (e.g., Roelofs & Piai, 2017) or is restricted to a small set of very long naming times (Scaltritti et al., 2015). Notably, these studies do not interpret their findings in the light of the temporal account but often assume that the link between experimental effects and naming times is mediated by attention. The attention of participants is assumed to fluctuate over the course of the experiment, with trials in which attention is less efficient resulting in longer naming times and in increased interference effects (Scaltritti et al., 2015; Roelofs & Piai, 2017). Piai et al. (2012) also discuss a scenario where lapses of attention result in a more superficial treatment of the distractor and in turn, in weaker interference or facilitation effects. The hypothesis that the distributional properties of the semantic interference effect are related to the participants' attentional state echoes back to the literature on the role of attention in the Stroop paradigm, which we will review in the next section.

In the previous section, we discussed the role of synchronization in processing for target and distractor words. It is possible that the distributional properties of the semantic interference effect are a consequence of the relationship between experimental effects and synchronization. We come back to this issue below.

*Attentional control in the Stroop paradigm*

In the present study, and following others, we define attention (or attentional control) as the ability to manage thoughts and actions in order to achieve specific goals, a process that requires selecting and managing the competition between relevant and irrelevant stimuli (e.g., Hutchison, 2011, see also Balota & Faust, 2001). Attention is often examined in tasks where participants have to respond to a target stimulus and to do so efficiently, must suppress irrelevant information or responses (i.e., conflict tasks). The Stroop paradigm is often used for this purpose. Distributional analyses of Stroop effects (e.g., Heathcote et al., 1991; Stroop, 1935) suggest both an overall shift in the distribution for incongruent trials as well as a positive skew of that distribution, as a result of a small proportion of very slow responses. Several mechanisms have been described to explain this pattern. Kane and Engle (2003), for instance, described two attentional mechanisms. The first allows the speaker to maintain the goal of the task over time. Failures of this mechanism result in errors and error corrections (also termed *periodic goal neglect*). The latter are visible in a small proportion of very slow response times in the incongruent condition. Note that Van Maanen and Van Rijn (2008) discussed the increased semantic interference in very slow trials in the picture-word interference paradigm in similar terms. Accordingly, on a restricted number of trials, the distractor is wrongly selected as a response, an error that needs to be corrected so that the correct response can be selected. Such "errors" are expected to occur more frequently in the related condition, and to generate abnormally long naming times. The second mechanism described in Kane and Engle (2003) is *competition resolution*. This mechanism is active in all trials, and responsible for the overall shift in the distribution of response times in the incongruent condition. Attentional processes in the picture-word interference paradigm can be assumed to be similar in nature, although the predominance of one or the other mechanism can depend on the proportion of congruent responses (e.g., Kane & Engle, 2003; Hutchison, 2011). We note here that distributional patterns in conflict tasks have received several different explanations, many of which do not directly resort to attention mechanisms. We come back to this issue in the General Discussion.

The purpose of accounts of attentional control in the Stroop task is not to describe language production processes and these accounts do not specify which planning processes may be affected by attentional mechanisms. Roelofs (2003; see also Roelofs, 2008), by contrast, discusses an account of attentional processes in the context of the WEAVER++ model (Roelofs, 1992), a model of language production. This model has been partly motivated by picture-word interference effects and was also applied to experimental effects in the Stroop task (Roelofs, 2003). In the WEAVER++ model, word representations are selected via spreading activation, and this process is monitored by condition-action rules. The system keeps track of the goal of the current task (e.g., name a picture) and attentional control makes sure to selectively activate lexical representations that are in line with the goal. For instance, following the visual processing of a picture, attentional control enhances the activation of lexical information when the goal is to name a picture. In the picture-word interference paradigm, two types of attentional mechanisms ensure that the name of the picture is produced rather than the name of the distractor. Output control ensures that the activation of lexical information corresponding to the picture's name is enhanced, and input control suppresses the lexical information associated with the distractor word (see also Piai et al., 2012).

Crucially for our purposes, attentional control (i.e., ability to focus on or maintain the goal of the task) is known to vary across trials. Many studies, yet in a different literature, have investigated (and reported) fluctuations of attention over the course of an experiment. Studies have shown for instance that in cognitive tasks, participants report states in which their thoughts shift away from the task at hand (i.e., mind-wandering states, e.g., Barnett & Kaufman, 2020; Smallwood & Schooler, 2006; Unsworth & Robison, 2016). These mind-wandering states have been related to EEG oscillatory patterns in the alpha frequency range (8-12 Hz). For instance, Compton et al. (2019, see also Arnau et al., 2020) reported that power in this range was higher preceding trials for which participants later reported being in a mind-wandering state. Variation in alpha power in the pre-stimulus period has also been related to response times in cued visuo-spatial tasks, with an increase in alpha power with increased response times (e.g., Kelly et al., 2009; Thut et al., 2006). Building on the latter findings, Jongman et al. (2020) examined the links between alpha power and response times in a simple picture naming task. They observed that higher alpha power in the motor system prior to stimulus onset or prior to speech onset was associated with *faster* (rather than slower) speech onset latencies.

*The current study*

In the present study, we report a series of analyses that aim at better qualifying the link between interference and facilitation effects, processing times, and attention, defined as a general mechanism that allows the speaker to focus on and maintain the goal of the task (i.e., name the picture). We seek evidence supporting the hypothesis that the size of experimental effects in the picture-word interference paradigm depends on the relative timing of processing for the distractor and target words (*synchronization hypothesis*). The distractor word only interferes when the information that it activates is available at the time when this information is relevant for the planning of the target word. We examine the distributional patterns of experimental effects and whether these patterns are as expected under the synchronization hypothesis. Finally, we examine the extent to which the participant's attentional state prior to stimulus presentation (indexed by EEG power in the alpha frequency range) modulates interference and facilitation effects. Addressing these issues will help better understand the mechanisms underlying the picture-word interference paradigm and highlight sources of variability in participants' behavior (within and across studies). Previous studies suggest that processing times and attention matter. Our aim is to verify these hypotheses and shed a novel light on how they contribute to observed experimental effects and their distributional properties. Note that our aim is not to describe the exact mechanisms by which attentional control operates to modulate experimental effects. We assume that on trials where participants are more focused, the efficiency of any attentional mechanism that contributes to ensure that the goal of the task is satisfied is enhanced. In the context of the Weaver++ model, for instance, in trials in which participants are more focused, both the input and output attentional mechanisms are likely to be more efficient.

Participants performed a picture-word interference naming task, with five conditions: no distractor (baseline), a phonologically related distractor, a semantically related distractor and two lists of unrelated distractors. After the picture-word interference task, the participants read all distractor words aloud. In order for the statistical power to be higher than in most previous picture-word interference studies (see Bürki et al., 2020 for an estimation of a posteriori power of several of these studies), 45 participants and 90 target pictures per condition were included.

In a first analysis, we examine interactions between the phonological facilitation and the semantic interference effects and a measure of difference in processing times for target and distractor words. This analysis tests the hypothesis that experimental effects occur for a restricted range of trials, those where the relevant information triggered by the processing of the distractor is active when this information is relevant for the planning of the target word. This synchronization hypothesis predicts that the two effects will interact either linearly or in a quadratic fashion with the difference measure. Moreover, given the assumption that semantic interference and phonological facilitation arise at different processing stages (lexical access or immediately preceding the articulation for the semantic interference effect, activation of phonological information for the phonological facilitation effect), this hypothesis further predicts that the range of values of the difference measure where the effect is maximal will differ between the two effects.

Next, we examine the distributional properties of the semantic interference effect. Past studies all report that the semantic interference effect increases as naming times increase, but diverge as to whether the effect is present over the entire response time distribution or restricted to very slow naming times. One aim of this first analysis is to clarify this issue with additional data. Moreover, the distributional properties of the semantic interference effect and that of the general interference and phonological facilitation effects are compared. To our knowledge, no study has yet looked at the distributional patterns of these two effects (a comparison between trials with unrelated distractors and trials without distractors was included in Roelofs & Piai, 2017, but was not analyzed, see Roelofs, 2008, Experiment 3, for a distributional analysis of phonological *interference* and, Experiments 1-2 for a distributional analysis of phonological facilitation in a picture-picture interference task).

In the literature on conflict tasks, distributional patterns are often taken to inform on the cognitive mechanisms underlying the participants' behavior in the task (e.g., De Jong et al., 1994; Spieler et al., 2000). Specifically, distributional analyses of response times examine whether an experimental manipulation leads to an overall shift in the distribution (as found for automatic semantic priming, Balota et al., 2008, or a "delayed start" as in persisting inhibition from a previous trial in negative priming, Tse et al., 2011), a shift of the distribution occurring together with a skew in which the interference increases across the distribution (as found for most conflict tasks) or decreases (as found for instance for the Simon task or a skew-only pattern (as expected if the experimental manipulation is entirely due to attention lapses, see De Jong et al., 1999). Notably and as explained and illustrated for instance in Pratte et al. (2010, see also Balota & Yap, 2011), the time course of an effect across the response time distribution is directly related to the distribution of response times in each of the conditions being compared. When the distributions of the response times in congruent and incongruent conditions only differ in their location (mean, i.e., there is a shift to the right for the incongruent condition but no skew), the effect (slower responses to the incongruent condition) is constant across the response time distribution[1]. When the incongruent condition also has a higher variance (i.e., shift of the distribution as well as skew for this condition), the effect increases with response times. When, by contrast, the fastest condition has a higher variance (i.e., shift of the distribution to the right and skew for this condition) the effect decreases with response times (see Figure 1 in Pratte et al., 2010). Under the hypothesis that experimental effects in the picture-word interference paradigm depend on the relative timing of processing between distractor and target word, we expect to see more variance in related conditions than in unrelated conditions. In related conditions, the impact of the distractor depends on whether the relevant information was activated at the right time. Response times for trials for which the relevant information was activated too early or too late will resemble response times for trials in the unrelated condition. These patterns in turn predict an increase in the semantic interference effect with increasing response latencies and a decrease in phonological facilitation as response latencies increase. By contrast, if an effect, e.g., the semantic interference effect is entirely due to lapses of attention, we expect a skew for the incongruent condition but no shift.

---

[1] As highlighted by Pratte et al. (2010) this reasoning holds when no part of the distribution in the slower condition has smaller values than the distribution in the fastest condition (or vice versa).

Finally, we examine the relationship between EEG power activity in the alpha band and experimental effects. We test whether the magnitude of interference and phonological facilitation effects can be predicted by differences in oscillatory patterns in the alpha band, prior to stimulus onset. We hypothesize that as their minds wander away from the task, participants will be less good at maintaining the goal of the task, that is naming the picture, which is likely to result in more interference from distractor stimuli. Whether the phonological facilitation effect should also be modulated by the participant's attentional state is unclear, and depends on whether attention only influences experimental effects via its impact on processing times, or also influences the activation/suppression of relevant information. Note that the demonstration that alpha power before stimulus onset influences interference effects would show that the ability to focus on the task matters, but would not inform on the specific attentional mechanisms involved.

**Method**

*Participants*

Forty-five participants, aged between 18 and 30 (mean = 23.2, SD = 3.5) took part in the experiment. They were all native German speakers. None of them reported hearing, psychiatric, neurological, or linguistic disorders. Their participation was rewarded either by course credit or money (48 € per participant). Participants were given details about the experimental procedure and provided their informed consent before the experimental session. The study received ethical approval from the Ethics committee of the University of Potsdam.

*Material*

We selected 90 German nouns (i.e., target words) and their corresponding pictures in the Multipic database (see Appendix 1; Duñabeitia et al., 2018). Eight additional pictures were selected to be used as filler and training items. Lemma frequencies for these words, according to the *dlexDB* database (Heister et al., 2011), ranged from 9 to 21184 occurrences per million (mean = 1991, SD = 3678). One hundred eighty German nouns were further selected to be used as distractors. Four hundred and fifty stimuli were created by combining the 90 pictures with the distractor words or a line of Xs, so as to form five conditions: (1) Baseline: A sequence of six Xs was superimposed on each of the 90 pictures; (2) Semantically related: Each picture was combined with a distractor word from the same semantic category (e.g., animals, tools); (3) Semantically unrelated: This condition was created by assigning each of the 90 distractors from the semantically related condition to a different picture such that picture and distractor words had no semantic or phonological relationship; (4) Phonologically related condition: Each picture was combined with a word overlapping in phonology with the target word (between one and four shared phonemes at word onset, mean = 2.6, SD = 0.7); (5) Phonologically unrelated condition: This condition was created by assigning each of the 90 distractors from the phonologically related condition to a different picture such that picture and distractor words had no semantic or phonological relationship. Sixteen additional distractor words were selected and combined with the filler pictures to form two trials from each condition. Table 1 below presents a summary of the properties of the target and distractor words.

*Table 1. Properties of the target and distractor words. For the latter, statistics are provided separately for the list used in the semantic vs. phonological contrast.*

| | Target words | | | |
|---|---|---|---|---|
| | Mean | SD | Min | Max |
| Lemma frequency | 1227 | 2336 | 1 | 12151 |
| Number of syllables | 2.14 | 0.82 | 1 | 5 |

| | Distractor words | | | | | | | |
|---|---|---|---|---|---|---|---|---|
| | Mean | | SD | | Min | | Max | |
| | List 1 | List 2 | List 1 | List 2 | List 1 | List 2 | List 1 | List 2 |
| Lemma frequency | 2480 | 1708 | 5195 | 3994 | 13 | 1 | 34023 | 26889 |
| Number of syllables | 2.08 | 2.03 | 0.74 | 0.84 | 1 | 1 | 5 | 4 |
| Number of letters | 5.72 | 5.82 | 1.44 | 1.75 | 3 | 3 | 11 | 12 |
| OLD | 1.96 | 2.07 | 0.52 | 0.742 | 1 | 1 | 3.55 | 4.2 |

List 1: List used for phonological contrast; List 2: List used for semantic contrast; OLD: Orthographic Levensthein Distance (see Yarkoni et al., 2008).

The text was displayed in white Arial font with a black outline. The first letter was an uppercase, all other were lowercases. Distractor-word pictures were presented within a gray square (see Figure 1 for examples).

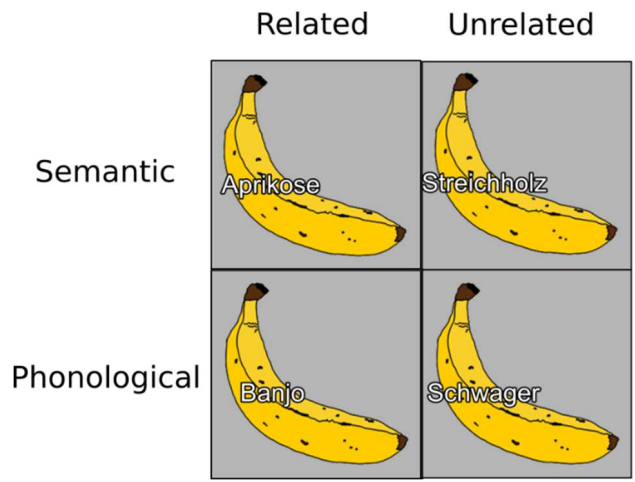

*Figure 1. Example of stimuli. The target word Banane 'banana' with a semantically related distractor, a phonologically related distractor, and two unrelated distractors.*

*Procedure*

Participants received written instructions at the beginning of the experimental session. The experimental session started with a familiarization phase. The 90 test pictures and eight filler pictures were presented one by one on the screen, without superimposed distractors but with the name of the picture written below. Participants were instructed to look at the pictures and to read their corresponding names. The task was self-paced. The order of presentation of the pictures was randomized.

The familiarization phase was followed by the picture-word interference task. In this task, the pictures together with a superimposed distractor or line of Xs appeared one by one on the screen, participants were instructed to name each picture as fast and accurately as possible. The task started with a short training session.

There were five experimental blocks. Each block started with the presentation of eight fillers followed by the presentation of 90 experimental trials. The presentation of the items was randomized for each participant with the following constraints: (1) each target word appeared once and in a different condition in each experimental block, (2) there was the same number of trials (18) from each experimental condition in each block, (3) over the whole experiment, half the items occurred in the phonologically related condition before occurring in the phonological unrelated condition and vice versa for the other half, and (4) half the items occurred in the semantically related condition before occurring in the semantically unrelated conditions and vice versa for the other half.

A trial started with the presentation of a fixation cross, whose duration ranged between 2200 and 2300 ms. The picture-word stimulus was then displayed in the center of the screen for 2300 ms. The visual angle of the square was 5.5 degrees and the vertical visual angle of the text was of about 0.6 degrees. The background color was grey. Participants had to provide their response before the disappearance of the picture (within 2300 ms). Vocal responses were recorded from stimulus onset to 3000 ms after picture onset. The Electroencephalographic (EEG) signal of the participants was monitored during the task.

After the picture-word interference task, the participants performed a delayed naming task (not detailed here) and a word naming (reading aloud) task. In the latter, participants were presented with the 180 words used as distractors in the picture-word interference task and were asked to read them aloud, as accurately and quickly as possible. At the onset of each trial, a fixation cross was displayed within a gray square located at the center of the screen for 700 ms. The written word then appeared, written in white Arial font with a black outline. The visual angle of the square was 5.5 degree (same spatial location as in the picture-word interference task). The word stayed on the screen for 1500 ms. The word was followed by an inter-stimulus interval of between 1200 and 2000 ms. The vocal response was recorded. The experiment started with four filler trials and entailed 180 test trials. The items were presented in a different random order for each participant.

**EEG acquisition and preprocessing**

Continuous EEG (sampling rate: 1000 Hz) was recorded with a BrainAmp MR amplifier, with 61 Ag/Ag-Cl electrodes, positioned according to the extended 10-system. FCz was used as the reference electrode during recordings. The EEGLAB toolbox (Delorme & Makeig, 2004) in MATLAB was used for preprocessing. The continuous EEG signal was first re-sampled at 500 Hz and recomputed against the average of all electrodes. A 0.1 Hz high pass filter (Kaiser windowed sinc FIR filter, order = 8008, beta = 4.9898, transition bandwidth = 0.2 Hz) and a 40 Hz low pass filter (Kaiser windowed sinc FIR filter, order = 162, beta =4.9898, transition bandwidth = 10 Hz) were then applied.

A first preprocessing pipeline was applied to correct the signal for visual artefacts. Epochs from -2s to 3.4s around picture onset were extracted and manually inspected. Noisy electrodes were spherically interpolated (mean number of interpolated electrodes per participant = 0.7, range = [0-5]), and noisy epochs excluded (mean number of rejected epochs per participant = 10.5, range = [0-47]). Artefacts corresponding to blinks were isolated using independent components analysis (ICA, Chaumon et al., 2015). To improve the decomposition, ICA was computed on the data sets filtered by a 1Hz high pass filter (Kaiser windowed sinc FIR filter, order = 802, beta = 4.9898, transition bandwidth = 2 Hz) and 40 Hz (Kaiser windowed sinc FIR filter, order = 162, beta =4.9898, transition bandwidth = 10 Hz). Resulting demixing matrices were then applied to the 0.1 Hz-40hz filtered data sets (for the use of a similar procedure, see Winkler et al., 2015). Components corresponding to blinking artefacts were then identified and excluded using both the SASICA plugin (Chaumon et al., 2015) and manual inspection.

The following pre-processing pipeline was then applied to the dataset. The whole pre-processing pipeline, for each participant, was applied and decided upon before any descriptive or inferential statistics were conducted. For each epoch, channels with an amplitude above +100 µV or below -100 µV were detected. Channels exceeding these thresholds on more than 45 epochs were spherically interpolated on each epoch. Epochs with three or more channels exceeding these thresholds were then disregarded. In epochs with one or two channels exceeding these thresholds, these channels were interpolated. A similar procedure was used with abnormal trends (*rejtrend* function; slope > 50 µV with R2 > 0.3). At the end of this procedure, each epoch was manually checked, noisy channels or epochs were excluded. Finally, we excluded epochs corresponding to trials with a production error and to trials with vocal onsets starting before 600ms or after 2300ms relative to picture onset. A mean of 90 epochs by participant (range = [16-219]) were excluded (i.e., 20% of trials).

Time frequency decomposition was performed by convolving EEG signals with a set of complex Morlet wavelets, defined as complex sine waves tapered by a Gaussian (Cohen, 2014). The set consisted in 55 complex Morlet wavelets, ranging from 3Hz to 30 Hz and resulting in linearly spaced steps of 0.5Hz, with a number of cycles ranging from 4 to 7 in linearly spaced steps.

We then computed a measure of power activity in the alpha frequency band (7.5 Hz to 12.5 Hz) in the time period ranging from -700 ms to -200 ms before picture onset. Power was computed for each epoch for a set of centro-temporo-occipital electrodes (FCz, FC1, FC2, FC3, FC4, FC5, FC6, FT7, FT8, Cz, C1, C2, C3, C4, C5, C6, T7, T8, CPz, CP1, CP2, CP3, CP4, CP5, CP6, TP7, TP8, Pz, P1, P2, P3, P4, P5, P6, P7, P8, POz, PO3, PO4, PO7, PO8, PO9, PO10, Oz, O1, O2). Electrodes were selected following Arnau et al. (2020). Frequencies in the alpha band were normalized (see Cohen, 2014)[2], i.e., expressed as a deviation from a baseline activity. Baseline activity was computed by taking the mean power across all time points in the pre-stimulus period, for each electrode and each frequency in the alpha band. For each analysis, the baseline was computed on the set of trials included in the analysis (e.g., in the analysis testing the effect of the position of the trial in the experiment on alpha power for pictures without distractors, the baseline was computed on trials without distractors, in the analysis testing the effect of power in the alpha band on response latencies and the interactions between power and experimental contrasts, only trials were included to compute the baseline). Note however that the analyses give the same results irrespective of whether the baseline includes all trials or only the trials included in the analysis. We then computed the power percent change for each frequency *f* in the alpha band, each electrode *e*, and each time point *t* in a given epoch following:

$$PercentChange_{f,e,t} = \frac{Power_{e,t,f} - Baseline}{Baseline}$$

---

[2] Time-frequency power decreases nonlinearly with increasing frequency (see Cohen, 2014, Chapter 18), with power at lower frequencies displaying larger magnitude than power at higher frequencies. For this reason, the mean of the raw power is biased towards lower frequencies. With a normalization procedure, all frequencies in the alpha band have the same weight.

We then averaged, for each epoch, the percentage change values of all frequencies within the alpha band and all electrodes.

Before each analysis, the dependent variable was screened for extreme values. These were defined based on the visual inspection of the distributions and were disregarded before conducting the inferential analyses. The raw and preprocessed EEG files are publicly available and can be found on OSF (https://osf.io/u46hf/)

*Analyses*

All analyses were conducted with the software R (R development core team, 2019). Scripts to reproduce the analyses are available on OSF (https://osf.io/u46hf/). All the statistical models that are reported are mixed-effects models and were conducted using the lme4 package Bates et al., 2015). Contrasts were determined manually (see Schad et al., 2020) such that the intercept represents the baseline, and four contrasts represent respectively the difference between phonologically related and unrelated trials (i.e., phonological contrast), the difference between semantically related and unrelated trials (semantic contrast), the difference between trials in the phonologically unrelated and baseline conditions (general interference contrast 1) and the difference between trials in the semantically unrelated and baseline conditions (general interference contrast 2). Related trials were coded as +1 and unrelated trials as -1. Trials in the baseline condition were coded as 1. Unless stated otherwise, the default random effect structure involved by participant and by item intercepts as well as by participant and by item slopes for the predictors of interest (when supported by the design). The random effect structure does not include random slopes for covariates or correlations between intercepts and slopes. The model was simplified in case of serious convergence issues, or if one of the variance components was equal or close to zero. Additional details and deviations from this default strategy are provided in the *Results and Discussion* section below. Each model was run twice, the first time on the whole dataset, the second time on the dataset trimmed of data points corresponding to residuals above an absolute value of 2.5 standard deviations in the first model (Baayen, 2008). The output of the second model is reported.

The first analysis was conducted to replicate the general interference, semantic interference, and phonological facilitation effects. In the second analysis, we examined the interactions between the semantic interference and phonological facilitation effects and a measure of relative processing speed between target and distractor words. To obtain this measure, we subtracted, for each target-word trial, the naming time for the distractor in the reading aloud task from the naming time in the baseline condition of the picture-word interference task for the target picture. In the third analysis, we organized the data in quantiles so as to examine how experimental effects change with quantiles (see for instance Ratcliff, 1979). In the last analysis, we tested the interactions between experimental effects and a measure of within participant variability in attention. In all analyses, we used the untransformed naming times as the dependent variable[3]. We use traditional criteria to decide whether a result should retain our attention or not (alpha = 0.05) and apply the Bonferroni correction when more than one model is performed to answer a given question. More details on each of these analyses are provided in the *Results and Discussion* section below.

**Results and Discussion**

Vocal responses were checked for accuracy and the time intervals between the onset of the picture presentation and the vocal response (i.e., response or naming latencies) for the correct responses were manually set based on the spectrogram and oscillogram using the *Praat* software (Boersma & Weenink, 2014). The resulting dataset can be accessed here (https://osf.io/u46hf/).

---

[3] Initially we had log transformed the latencies in all but the distributional analysis. Following the concern formulated by a reviewer we re-ran all models with the untransformed latencies as the dependent variable. None of the outcome is affected by this change of procedure.

There were 19090 correct responses out of 20250 trials (94%). The vocal response was uncertain for seven trials, these trials were disregarded. Most errors resulted from the use of a different word than the intended word (766 out of 1153 errors, 66 %). The remaining errors were dysfluencies (N = 231) and no responses (N = 156). Trials with errors as well as one additional trial where the participant could be heard yawning were disregarded. Given the small percentage of errors, no analysis was conducted on these errors. The 46 data points for which the participants' responses started after the disappearance of the picture (2300 ms after picture onset) were disregarded. The remaining 19043 data points were included in the analyses. The mean naming latency was 876 ms (SD = 255), with mean naming latencies ranging from 708 ms to 1240 ms across participants. Figure 2 below displays the mean response latencies in each condition.

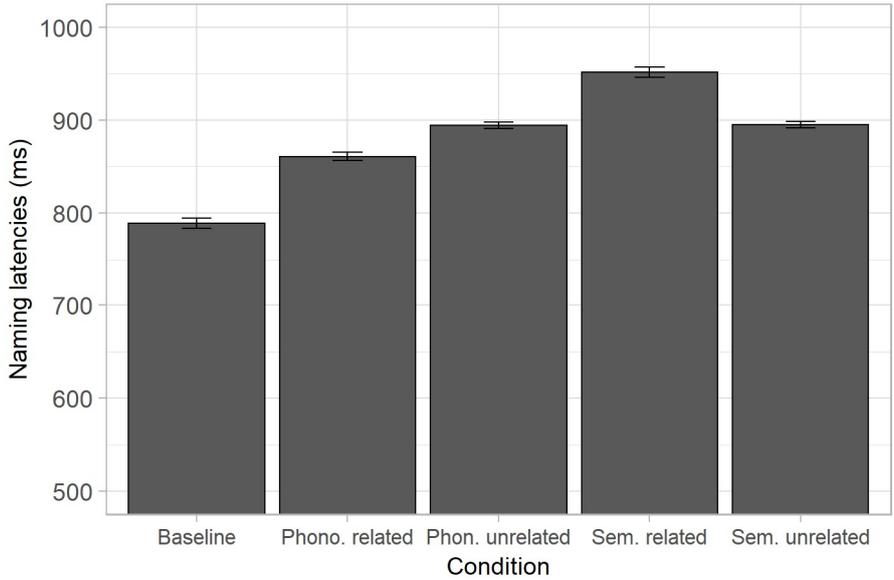

*Figure 2. Observed mean naming latencies and standard errors (values are adjusted for within-Participant designs following Morey, 2008) for each condition.*

*Analysis 1. Replication of picture-word interference effects*

The first statistical model was conducted to replicate the general interference effect, phonological facilitation effect, and semantic interference effect. The four experimental contrasts were entered in the analysis as fixed-effects. To account for the effect of repetition of the target word and effects of habituation or fatigue, we also included, as covariates, the number of repetitions of the target word and the position of the trial in the experiment. These two variables were centered around the mean. The results of this analysis are presented in Table 2.

Table 2. Output of mixed-effects model with the four experimental contrasts, Number of repetitions, and Position of the trial in the experiment as fixed-effects (Analysis 1). The intercept represents the naming latencies in the Baseline condition.

| Predictor | Estimate | 95% Conf. Int. | p |
|---|---|---|---|
| (Intercept) | 772.7 | 743.0 , 802.5 | <0.001 |
| General interference contrast (list 1) | 97.9 | 89.6 , 106.2 | <0.001 |
| General interference contrast (list 2) | 98.0 | 88.1 , 107.9 | <0.001 |
| Phonological contrast | -35.8 | -49.6 , -22.0 | <0.001 |
| Semantic contrast | 53.0 | 37.5 , 68.5 | <0.001 |
| Number of Repetitions (centered) | -58.3 | -66.5 , -50.2 | <0.001 |
| Position of trial in experiment (centered) | 0.37 | 0.28 , 0.46 | <0.001 |

This first analysis supports the hypotheses that when compared to unrelated distractor words, phonologically related distractors create facilitation and semantic distractors create interference. Moreover, unrelated words create interference when compared to a picture without distractor. We further note that the naming latencies decrease with the repetition of the target word and increase with the position of the trial in the experiment. It is also worth noting that the naming times for the two unrelated conditions are almost numerically identical. A visual representation of the naming time distributions for the different conditions is presented in Appendix 2. As can be seen, the two distributions are almost identical. This suggests that the two unrelated lists do not differ in important aspects. The results of a statistical model with the two-ways interactions between experimental contrasts and number of repetitions as well as position of the trial in the experiment is presented in Appendix 3.

*Analysis 2. Differences in processing times for distractor and target words*

In the next analysis, we examined whether experimental effects in the picture-word interference paradigm are modulated by the difference in processing times for the target and distractor words. To obtain a measure of difference in processing times between target and distractor words, we computed, for each trial, the difference between the naming time for the same picture in the baseline condition and the reading time for the distractor used in this trial. In the reading aloud task, participants made 308 errors (N = 8100, about 4%). We disregarded these data points and further disregarded 33 data points for which the annotator was not certain that the word had been read correctly / could not set the onset precisely, and six data points with naming times below 100 ms. The mean reading times for the remaining 7752 data points was 591 ms (SD = 140). Reading times ranged between 294 ms and 1562 ms across trials and between 426 ms and 838 ms across participants. On average, participants responded to the picture about 275 ms slower than to the distractor.

The difference measure was centered around the mean. We tested the interactions between this variable (linear and quadratic terms) and the phonological and semantic contrasts. The models had random intercepts and slopes for all main effects (the results of a model with random slopes for all main effects and interactions show the same pattern). The output of this analysis is presented in Table 3 below, and illustrated in Figure 3.

*Table 3. Output of linear mixed-effects model testing for interactions between the phonological and semantic contrasts and the difference in processing times for the target and distractor words used in each trial (Analysis 3).*

| *Predictor* | *Estimate* | *95% Conf. Int.* | *p* |
| --- | --- | --- | --- |
| (Intercept) | 866.8 | 819.4 , 914.3 | <0.001 |
| Phonological contrast | -35.3 | -47.2 , -23.3 | <0.001 |
| Difference measure (linear) | 1784.9 | 1151.0 , 2418.8 | <0.001 |
| Difference measure (quadratic) | 662.2 | 144.9 , 1179.5 | 0.012 |
| Semantic contrast | 50.7 | 43.7 , 57.6 | <0.001 |
| Semantic contrast * Difference measure (linear) | -743.0 | -1563.2 , 77.2 | 0.076 |
| Semantic contrast * Difference measure (quadratic) | -1357.6 | -2206.4 , -508.9 | 0.002 |
| Phonological contrast * Difference measure (linear) | 234.5 | -597.3 , 1066.3 | 0.581 |
| Phonological contrast * Difference measure (quadratic) | -1156.9 | -1974.7 , -339.1 | 0.006 |

I

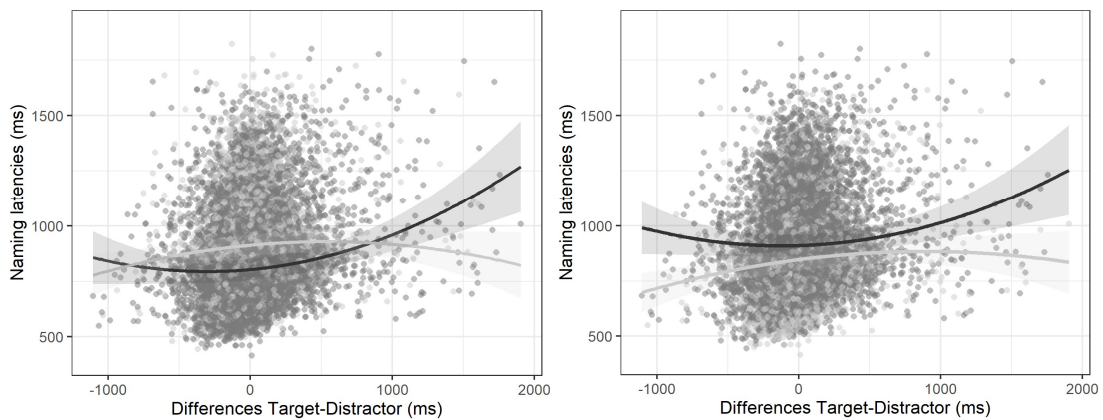

*Figure 3. Interactions between the difference measure and the semantic contrast (left) and between the difference measure and the phonological contrast (right) as predicted by the statistical model. Black line = unrelated trials, Grey line = related trials.*

This analysis reveals that both contrasts interact with the quadratic term of the difference measure. The semantic contrast is found when the difference between naming and reading is around its mean, at about 275ms. The interaction further suggests that if the distractor is processed too quickly relatively to the planning of the target word (i.e., large value for the difference measure), the semantic interference effect reverses to a semantic facilitation effect. The phonological contrast on the other hand is minimal when the semantic interference effect is maximal. It increases when the difference measure increases that is, presumably when the distractor word is processed more quickly relative to the target word. The plot suggests that it might also decrease when the difference measure decreases (i.e., when the distractor word is processed more slowly relative to the target word) but here less data points are concerned). Interestingly, whereas the semantic interference effect is maximal for most data points, the phonological facilitation effect is maximal only for a restricted set of data points. This could explain the difference in effect sizes between these two effects, as well as differences in effect sizes across studies (in the present study, the estimate of the semantic interference effect is particularly high, when compared to other studies, see for instance Bürki et al., 2020).

*Analysis 3. Distributional analyses*

We performed distributional analyses of the four experimental effects. For each participant, individual trials were organized in ascending order and divided into five quantiles (e.g., Ratcliff, 1979). Figure 4 below displays the mean for each condition in each quintile.

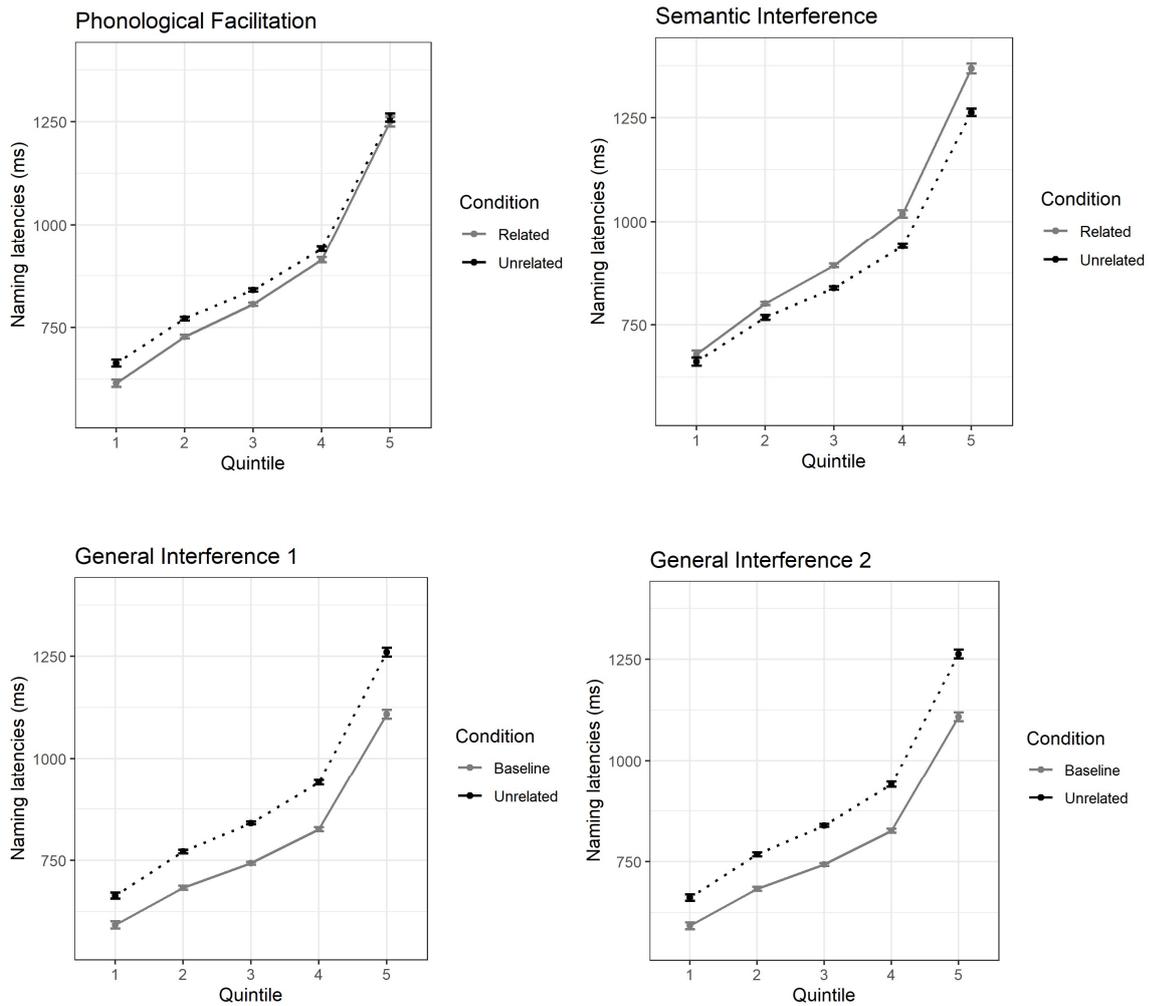

*Figure 4. Mean in related and unrelated conditions for the phonological contrast (upper left panel), semantic contrast (upper right panel), and the two general interference effects (lower panels) in each by participant quintile. Error bars represent standard errors, adjusted for within participant designs*

As can be seen on the graphs, the difference between the phonologically related and unrelated conditions decreases with quintile. The reverse is observed for the semantic interference and general interference effects, although the increases appear smaller for the latter effects. Moreover, the graphs also suggest that none of these effects seems to be restricted to the slowest part of the distribution[4].

---

[4] Two approaches have been used to investigate the distributional pattern of the semantic interference effect, vincentization and ex-Gaussian analyses. In the present contribution, we opted for vincentizations. Ex-gaussian

We tested these trends in statistical models. We conducted one mixed-effects model for each experimental contrast and tested the interaction between this contrast and quintile. We used 1$^{st}$, 2$^{nd}$ and 3$^{rd}$ degree polynomial contrasts for the variable quintile. Repetition was entered as a covariate in the models. Given that we conducted four different models, (a model with all contrasts and their interactions resulted in too many parameters) we applied the Bonferonni correction. An effect or interaction is considered significant if the corresponding *p*-value is below 0.05/4. All models had by participant and by item random intercepts and random slopes for all main effects). The results of these models are presented in Appendix 4 and naming times in each quintile, as predicted by these models, are displayed in Appendix 5. In all models, the linear, quadratic, and cubic terms for the variable quintile are significant and the linear term for the interaction is significant.

In the model testing the interaction between the phonological contrast and quintile, the facilitation effect decreases linearly with quintile. In the model testing for the interaction between the semantic contrast and quintile, the linear and quadratic terms for the interaction are significant, indicating that the interference increases linearly with quintile, and that this increase is more pronounced in the first quintiles. Finally, the models testing the interactions between quintile and the two general interference effects show a linear increase in general interference with quintile, with a quadratic trend for one of these effects.

We also conducted one model with all four experimental contrasts for each quintile. We corrected for multiple comparisons using the Bonferroni correction. Effects are considered significant at 0.01 (0.05/5). The details of these analyses can be found in Appendix 6. General interference and semantic interference effects are found in all quintiles, the phonological facilitation effect in all but the last quintile. These analyses provide support for the hypotheses that the semantic and general interference effects increase linearly with quintile while the phonological facilitation effect decreases linearly with quintile. Moreover, they show that the relationship is not merely linear, with a quadratic or cubic

---

analyses are very sensitive to both outliers and removal of outliers. Moreover, the descriptive statistics in the present study clearly show that the effects are not restricted to the tail of the distribution.

component depending on the contrast. In this analysis, the vincentization was performed by participant. It therefore provides information on changes in experimental effects that occur as a given participant responds slower or faster. The interactions between experimental effects and quintiles therefore reflect variability in trials or target words. Next, we performed the same analyses but organized the data in quintiles per item. That is, for each target word, individual trials were organized in ascending order and divided into five quantiles. A visual representation of the mean in each condition in each quintile and for each experimental contrast is displayed in Figure 5.

The statistical results are displayed in Appendix 7 and Appendix 8. They are similar to the results of the by participant distributional analysis. All four experimental effects interact linearly with the variable quintile, they are present in all quintiles except for the phonological facilitation, absent in the last quintile.

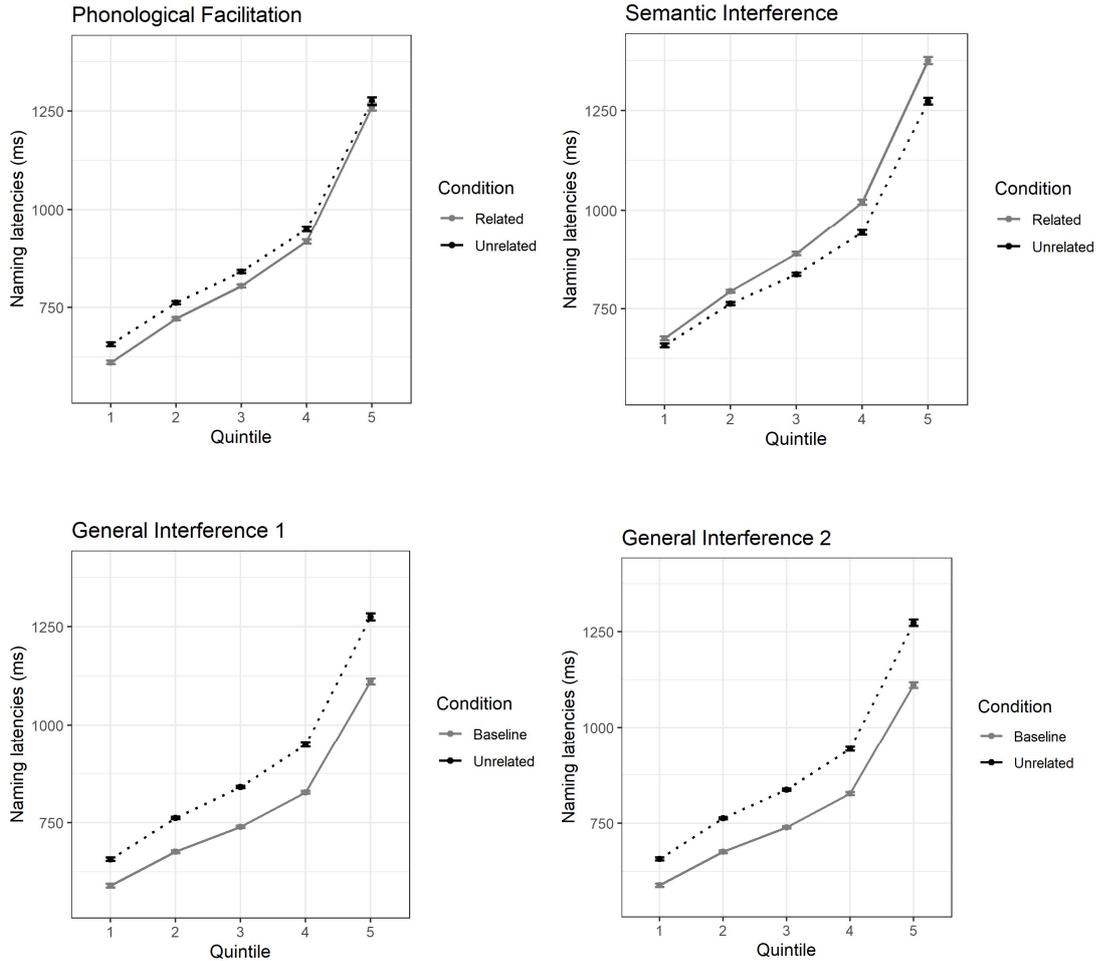

*Figure 5. Mean in related and unrelated conditions for the phonological contrast (upper left panel), semantic contrast (upper right panel), and the two general interference effects (lower panels) in each by item quintile. Errors bars represent Standard Errors, adjusted for within participant designs*

Such distributional patterns for the semantic and general interference effects are expected if response times in the slow condition are more variable than response times in the fast condition. Likewise, the distributional pattern for the phonological facilitation is expected if the variance is larger in the faster condition (i.e., the related condition). Density plots of the response times in each condition can be found in Appendix 2. A higher variance in related conditions is for instance expected under the synchronization hypothesis. Under this hypothesis, the impact of the distractor in related trials depends on the relative timing of processing between target and distractor information – and will therefore vary across trials.

To summarize, we observe similar results whether we organize the quantiles as a function of participants or items. In both cases, the experimental effects depend on the distribution of the response latencies. Interference effects increase with quintile and the phonological facilitation effect decreases with quintile. The effects can be found in all quintiles, except for the phonological facilitation effect, which is absent in the last quintile. These distributional patterns are for instance expected under the hypothesis that experimental effects depend on the synchronization in processing for the target and distractor word. Notably however, they do not necessarily rule out alternative hypotheses. We come back to this issue in the General Discussion.

*Analysis 4. EEG power in alpha frequency band*

In the last analysis, we examined whether variations in EEG power in the alpha frequency band modulated experimental effects. As a sanity check, we first tested whether alpha power was predicted by the position of the trial in the experiment (centered around the mean). Recent experimental data have indeed suggested that power in the alpha band increases over the course of an experiment (Benwell et al., 2019). Other studies reported that alpha power in this frequency band was higher when participants reported being in mind-wandering states (e.g., Compton et al., 2019). Under the hypothesis that attention decreases over the course of the experiment, we would expect that power in the alpha frequency band increases over the course of the experiment. Alternatively, the relationship between attention and trial position could be quadratic.

We first restricted the dataset to the trials in the baseline condition and fitted a linear mixed-effects model with alpha power as the dependent variable, position of the trial (centered around the mean, linear and quadratic terms), and number of repetition of a given picture (also centered around the mean) as predictors. The results of this model is reported in Table 4. In this model, the random structure does not include a by item random slope. Results show that power in the alpha frequency band increases with the position of the trial in the experiment. The quadratic term for trial position was not significant and was removed from the final model. The same models were also run on all trials, with similar results. Interestingly, power in the alpha band decreases with repetition (or block). Note that the results remain the same when response times are entered in the analysis (response times do not significantly contribute to the model). We further note that alpha power is not equal across condition, with a lower power for the semantically related than unrelated condition.

*Table 4. Results of statistical models with alpha power as the dependent variable and position of the trial and repetition as fixed effects*

|  | Power percent change in alpha band | | |
| --- | --- | --- | --- |
|  | *Estimates* | *95% Conf. Int.* | *p* |
| (Intercept) | -8.11 | -9.51 , -6.71 | <0.001 |
| Trial position (centered) | 0.12 | 0.09 , 0.15 | <0.001 |
| Number of Repetitions (centered) | -8.38 | -10.65 , -6.11 | <0.001 |

Next we examined whether alpha power interacted with experimental conditions to predict naming times. We added alpha power to the model of Analysis 1. The variance for the by-item random slope was equal to 0, this random slope was removed from the model. The results of this model are presented in Table 5.

*Table 5. Results of statistical model predicting response latencies as a function of alpha power with interactions between alpha power and experimental contrasts*

|  | Naming latencies | | |
|---|---|---|---|
| *Predictors* | *Estimates* | *95% Conf. Int.* | *p* |
| (Intercept) | 773.07 | 742.35 – 803.80 | <0.001 |
| Alpha Power | 0.04 | -0.05 – 0.14 | 0.394 |
| General interference contrast (list 1) | 95.43 | 88.05 – 102.81 | <0.001 |
| General interference contrast (list 2) | 96.09 | 88.72 – 103.47 | <0.001 |
| phonological contrast | -34.87 | -42.30 – -27.43 | <0.001 |
| semantic contrast | 51.45 | 43.96 – 58.93 | <0.001 |
| Position of trial in experiment (centered) | 0.40 | 0.30 – 0.50 | <0.001 |
| Number of Repetitions (centered) | -60.91 | -69.25 – -52.57 | <0.001 |
| Alpha Power * General interference contrast (list 1) | 0.03 | -0.10 – 0.17 | 0.639 |
| Alpha Power * General interference contrast (list 2) | -0.03 | -0.17 – 0.11 | 0.664 |
| Alpha Power * phonological contrast | -0.08 | -0.22 – 0.06 | 0.270 |
| Alpha Power * semantic contrast | -0.00 | -0.14 – 0.14 | 0.962 |

To summarize, the impact of the position of the trial in the experiment and of repetition on EEG alpha power is in line with the hypothesis that EEG alpha power is somehow related to the participant attentional state. Our analyses do not offer support for (or against) the hypothesis that power in the alpha band modulates experimental contrasts. There is also no statistical support for the hypothesis that power in the alpha frequency band modulates naming latencies. In order to obtain additional information on the main effect of EEG alpha power and its interaction with each of the experimental contrasts, we conducted Bayes Factors. These analyses were conducted with the package brms (Bürkner, 2017). For each contrast, we computed four Bayes Factors. The first two compared the model with alpha power, the experimental contrast, and their interaction, to the same model without the interaction. The third and fourth Bayes Factors compared the latter model to a model without the main effect of alpha power. All models also had the position of the trial in the experiment as well as repetition as covariates, and random intercepts for participant and item. To assess the sensitivity of the Bayes Factors to the priors, each Bayes Factor was computed with two different priors for the main effect of alpha power and for the interaction. The first (prior 1) was a normal distribution, centered at 0 with a standard deviation of 10. The second (prior 2) was a normal distribution, centered at 0 with a standard deviation of 1. Whereas the first prior is rather uninformative (i.e., associated with a wide range of effect sizes, including large effect sizes), the second assumes a very small effect size. Priors for the other parameters of the model are detailed in Appendix 10. Resulting Bayes Factors are displayed in Table 6.

*Table 6. Bayes factors comparing models with main effect of EEG alpha power and its interactions with experimental contrasts. The Bayes Factors in the table quantify the evidence in favor of the simpler model (without the main effect or interaction).*

| $M_0$ | $M_1$ | $BF_{0,1}$ Prior 1 | $BF_{0,1}$ Prior 2 |
|---|---|---|---|
| EEG alpha power + Semantic contrast | EEG alpha power * Semantic contrasts | 80.9 | 11.4 |
| Semantic contrast* | EEG alpha power + Semantic contrasts | 153.8 | 15.9 |
| EEG alpha power + Phonological contrast | EEG alpha power * Phonological contrast | 99.9 | 10.8 |
| Phonological contrast | EEG alpha power + Phonological contrast | 154.7 | 14.7 |
| EEG alpha power + General interference contrast 1 | EEG alpha power * General interference contrast 1 | 73.4 | 6.8 |
| General interference contrast 1 | EEG alpha power + General interference contrast 1 | 140.7 | 18.6 |
| EEG alpha power + General interference contrast 2 | EEG alpha power * General interference contrast 2 | 53.0 | 9.0 |
| General interference contrast 2 | EEG alpha power + General interference contrast 2 | 134.2 | 14.9 |

All Bayes Factors favor models without the main effect of EEG alpha power or the interaction between EEG alpha power and experimental contrasts. This is also true with priors that assume very small effect sizes.

**General Discussion**

The aim of the present study was to examine the relationships between several experimental effects in the picture-word interference paradigm, processing speed, and attention. We focused on three effects: the general interference effect, the semantic interference effect, and the phonological facilitation effect.

Our experiment successfully replicates interference and facilitation effects. We further observed interactions between the semantic interference and phonological facilitation effects and a measure accounting for the difference in processing times for target and distractor words. The relationship is quadratic, and shows that both effects occur in a restricted time range of the difference measure. This finding echoes back to the numerous studies showing that both the semantic interference and phonological facilitation effects depend on the relative timing of *presentation* of the distractor and target word (i.e., the SOA, e.g., Schriefers et al., 1990) and provides a more direct demonstration that the source of these interactions is indeed the relative timing of *processing* of the two words. Interestingly, we observe that the semantic interference effect can turn into facilitation when the distractor is processed quickly relative to the planning of the picture's name. This finding is consistent with several proposals in the literature. In many accounts of the semantic interference effect (Bloem & La Heij, 2003; Roelofs, 1992) the distractor word activates its associated concept and this concept in turn sends activation to related concepts, among them, the target word. Abdel Rahman and Melinger (2019)'s swinging lexical network for instance assumes facilitation at the conceptual level and interference at the lemma level. Piai et al. (2012) mention the possibility that net experimental effects of semantically related distractors can result from both facilitation and interference. Our finding suggests that facilitation dominates when the distractor is processed quickly relative to the planning of the target word. More generally, these findings are in line with the temporal hypothesis and the results reported by Geng et al. (2014), that experimental effects in the picture word interference task depend on naming times. Gheng et al. reported some evidence that this was the case for the distractor frequency effect, we extend this demonstration to interference and phonological facilitation effects.

As discussed in the introduction, Bürki (2017) also examined the interaction between phonological facilitation and a difference measure. In this study, the interaction was linear rather than quadratic. Moreover, an increase in the difference measure lead to a decrease in the facilitation effect. Given that naming times (to pictures and written words) differ greatly across pictures and words (e.g., as a result of differences in their lexical properties) the range of values at which the effect will be maximal cannot be expected to be the same across studies, when these use different stimuli. In Bürki (2017) there was no main effect of phonological facilitation when target and distractor overlapped in their first syllable, but there was such an effect for pairs that overlapped in their second syllable. This pattern of results suggests that in that study, for many trials, the phonological information corresponding to the onset of the distractor word was not active at the time where it could have been relevant for the preparation of the picture's name. We speculate that the "shape" of the interaction between the difference measure and experimental effects depends on the experimental material. To further shed light on this issue, we analyzed the data of another study, with the same design and material as in the experiment reported here, but different participants. The interactions for both the semantic and phonological contrasts show the same patterns. The outcome of this analysis is displayed in Appendix 9. It should further be noted that in Bürki (2017), and as discussed already, the naming and reading latencies used to compute the difference scores were from trials with pictures and distractors. In the present study, the naming latencies were collected for trials without distractors, and the reading latencies for trials without pictures. These differences may have further influenced the naming and reading latencies, as well as, as a consequence, the difference scores.

The demonstration that semantic interference and phonological facilitation effects depend on the relative timing of processing between target and distractor words has important methodological consequences. As discussed in the Introduction, many studies use these effects as markers of underlying processes (e.g., phonological facilitation as a marker of phonological encoding, semantic interference as a marker of lexical access, e.g., Meyer, 1996). For instance, several authors used the phonological facilitation effect to examine the scope of phonological advanced planning and show that this effect depends on the type of utterance (e.g., bare noun naming versus noun phrases, Klaus & Schriefers, 2018, adjective-noun versus noun-adjective, Michel Lange & Laganaro, 2014). In the light of the present study's findings, the question arises of whether the interactions between utterance type and phonological facilitation reported in these studies reflect the scope of advanced planning or are a mere consequence of differences in the temporal alignment between target and distractor word processing across utterance types.

The distributional analyses replicate the finding that the semantic interference effect depends on the response time distribution and further shed light on the nature of this distributional pattern. They confirm that the semantic interference effect can be found across the entire distribution of naming times and support the claim by Roelofs and Piai (2017) that the effect is not selectively linked to abnormally slow naming times. This result does not support the hypothesis that the semantic interference is restricted to the tail of the distribution either because it only surfaces in those trials were attention is operating the least efficiently (Scaltritti et al., 2015) or because it reflects a restricted set of trials where the participant selected and encoded the wrong word (i.e., the distractor), which then had to be inhibited and replaced by the target response. We note here that this distributional pattern (i.e., an effect size that increases with response times) is by far the most frequent distributional pattern in conflict tasks, as well as more generally, in tasks with strength manipulations (Pratte et al., 2010). In language production, Shao et al. (2014) reported for instance a similar pattern for the effect of naming agreement on picture naming latencies and we observed the same pattern in an unpublished experiment manipulating age of acquisition. Pictures with low name agreement or whose names were acquired later in life are named with slower and more variable naming times, and this effect increases with response times.

Our analyses further provide information on the distributional properties of two other effects, the general interference effect and the phonological facilitation effect. The general interference effect exhibits the same pattern as the semantic interference effect. The similarity in distributional patterns for the two interference effects is in line with the widely held assumption that the two effects are at least partly governed by the same underlying mechanisms. More generally, this pattern is in line with the default pattern in conflict tasks, where trials with conflicting information are slower and more variable. The reversed distributional pattern is observed for the phonological facilitation effect. This effect decreases with quintile, and is no longer present in the last quintile, where the two interference effects are maximal. To our knowledge, this is the first demonstration that phonological facilitation in the picture-word interference task is not constant across the naming time distribution. This pattern contrasts with Roelofs (2008) who found no interaction between the phonological contrast and decile. Importantly, however, phonological overlap in that study generated interference rather than facilitation. According to the author, the fact that the participants had to name the distractor on a subset of trials might have increased the competition of the distractor words in other trials.

The distributional patterns that we observe for the semantic interference and phonological facilitation effects are expected under the hypothesis that the synchronization in the processing of target and distractor words determines the extent of interference and facilitation effects for a given trial and fit well with the finding that these effects are modulated by the difference in processing times between target word and distractor word processing. This hypothesis predicts that related trials are more variable than unrelated trials, because not all trials are influenced by the linguistic properties of the distractor. As discussed in the Introduction (and for instance illustrated in Pratte et al., 2010), when an experimental condition is both more difficult and more variable than the other condition, the difference between the two will increase with the response time distribution. When, by contrast, an experimental condition is easier but more variable than the other, the difference between the two decreases with the response time distribution. Interestingly, Spieler et al. (2000) also discuss the distributional pattern of the Stroop effect in similar terms: "*One interpretation of the interference effect observed in τ is that the influence of the word dimension is not consistent across trials*" (p. 511).

Notably, whereas our findings are compatible with the hypothesis that the distributional patterns of experimental effects are at least partly caused by variations in the temporal alignment between target word and distractor word processing across the naming time distribution, our study does not provide a direct test of this hypothesis and does not rule out other (not necessarily mutually exclusive) accounts of these distributional patterns. In the next paragraphs, we discuss a subset of accounts of distributional patterns in conflict tasks. Note that a detailed review of these accounts and how they could explain phonological facilitation and semantic interference effects in the picture-word interference paradigm is beyond the scope of the present paper. A detailed discussion of how the different accounts of conflict tasks and their mechanisms (e.g., spreading activation, persisting inhibition, response competition, etc.) could explain phonological facilitation and semantic interference effects and their distributions would need to consider different assumptions about the functional origin of these effects and, more generally, about the processes involved in producing words. For instance, whereas some word production models assume that lexical access is competitive and involves inhibition (e.g., Abdel Rahman & Aristei, 2010; Levelt et al., 1999) other models assume no competition/inhibition. Moreover, whereas in the first the semantic interference effect is a direct consequence of lexical competition, in the latter, this effect arises during pre-articulatory processes. For detailed discussions in the context of the Weaver ++ model, we refer the reader to the work of Ardi Roelofs (e.g., Roelofs, 2003, see also Piai et al., 2013).

As mentioned above, many congruency effects (slower response times in the incongruent than in the congruent condition) in conflict tasks have been shown to increase with response times and most accounts of these effects/tasks can explain this default pattern (see also Pratte et al., 2010). Negative delta patterns (decreasing congruency effects with response times) have been reported in only a small set of tasks or conditions (e.g., Simon task) and are more challenging for these accounts. An oft cited account of negative delta patterns is the activation-suppression account (Ridderinkhof, 2002).

According to this account, response times in conflict tasks result from two processes. At first, the task-irrelevant stimulus is automatically activated. This automatic activation is immediately followed by the inhibition of the activated response. This second process is under executive control and builds up in a gradual way. Given that inhibition takes time to deploy, it is more efficient for longer responses. This model could account for the phonological facilitation effect and its distributional pattern. The related condition would be considered the congruent condition, and trials in this condition would benefit more from an early automatic activation of the distractor. The inhibition of the task irrelevant stimulus (distractor) would apply similarly to the two conditions in trial with slow responses. To the best of our knowledge, the activation-suppression account cannot explain the positive distributional pattern of interference effects. Some authors have suggested that selective inhibition could be applied in the semantic interference effect, at least by some participants (Shao et al., 2015). The increase in the semantic interference effect in the last two quintiles in the present study is at odds with the hypothesis that as a group, participants apply more inhibition in slow responses. Hence, the activation-suppression account does not seem to be able to provide a unified account of interference and facilitation effects.

An alternative explanation of negative delta plots is provided by Ulrich et al. (2015). Interestingly, their model assumes that distributional effects depend on the time course of activation of task-irrelevant information. In classical diffusion models (Ratcliff, 1978), information accumulates over time until one threshold is reached, which triggers the execution of the response. There are two possible thresholds in conflict tasks, the first corresponds to task relevant information (correct response) the second to task irrelevant information. According to Ulrich et al. (2015) diffusion models cannot explain negative delta patterns. The authors discuss an extension of these models, the *Diffusion Model for Conflict tasks*, where distributional properties depend on the time course of activation of task-irrelevant information. If this activation is maximal shortly after stimulus onset, negative slopes are expected. Our proposal that distributional patterns in the picture-word interference paradigm at least partly result from variability in the synchronization of processing between target and distractor words across trials seems to fit well with this proposal.

In our last series of analyses, we examined EEG power in the alpha frequency band, under the assumption that variability in this measure indexes fluctuations of attention. Power in the alpha band was found to increase with the position of the trial in the experiment and to decrease with repetition. The first effect is not unprecedented (see Benwell et al., 2019), and this replication suggests that alpha power captured variability in attention in the present experiment. To our knowledge, effects of repetition are not documented. We speculate that attention may be boosted by familiarity with the material. Having seen a familiar item might boost the participant's attention to the next item. Our analyses do not provide support for the hypothesis that power in the alpha band modulated experimental effects or even naming latencies. Rather, Bayes Factors suggest that EEG alpha power does not impact naming times, facilitation, or interference effects in the present experiment. Whereas this result is as expected under the hypothesis that attention does not modulate naming times or experimental effects, it remains an absence of effect and as such, must be considered with care. At least one study reported results that suggest a link between naming times and alpha power (Jongman et al., 2020) and additional studies are clearly needed to replicate this link and assess the conditions in which it can be observed.

**Conclusion**

Interference and facilitation effects in the picture-word interference paradigm depend on the synchronization in processing times between the picture and word stimuli. The distributional properties of these effects could reflect this relationship. These findings shed a novel light on the mechanisms underlying these effects and the factors that modulate them. Importantly, they show that variability in experimental effects across studies or conditions are to be expected as a result of mere differences in processing times. Whereas variability in synchronization between picture and word processing could potentially result from fluctuations of attention, using EEG alpha power prior to trial onset as an index of attention, we were not able to find evidence that attention modulates the participants' performance in the experiment reported here.

**Acknowledgements**

This research was funded by the Deutsche Forschungsgemeinschaft (DFG, German Research Foundation) – project number 317633480 – SFB 1287, Project B05 (A. Bürki). The authors would further like to thank Andreas Mädebach, an anonymous reviewer and Keith A. Hutchison. Their comments on previous versions of the manuscript were extremely helpful.

*Appendix 1. Material used in experiment*

|  | Distractor type | | | |
|---|---|---|---|---|
| **Target word** | **Semantically- related** | **Semantically- unrelated** | **Phonologically - related** | **Phonologically - unrelated** |
| Dusche (shower) | Badewanne (bathtub) | Frisbee (frisbee) | Duell (duel) | Banjo (banjo) |
| Knochen (bone) | Muskel (muscle) | Radar (radar) | Knospe (bud) | Salbe (ointment) |
| Leopard (leopard) | Tiger (tiger) | Kuchen (cake) | Lehm (clay) | Dimension (dimension) |
| Zwiebel (onion) | Knoblauch (garlic) | Stock (stick) | Zwiesel (zwiesel) | Antike (antiquity) |
| Mikroskop (microscope) | Fernglas (binoculars) | Strumpf (stocking) | Mikado (mikado) | Geschichte (story) |
| Spinne (spider) | Ameise (ant) | Duschgel (shower gel) | Spitzel (spy) | Duell (duel) |
| Kleeblatt (shamrock) | Schilf (reed) | Zange (pliers) | Klerus (clergy) | Hamster (hamster) |
| Bürste (brush) | Kamm (comb) | Rubin (ruby) | Bürde (burden) | Chaos (chaos) |
| Ellenbogen (elbow) | Knie (knee) | Postkarte (postcard) | Elfe (elf) | Feudalismus (feudalism) |
| Brief (letter) | Postkarte (postcard) | Wal (whale) | Brigade (brigade) | Taste (button) |
| Zeitung (newspaper) | Buch (book) | Herd (stove) | Zeichen (sign) | Trost (consolation) |
| Kerze (candle) | Fackel (torch) | Muskel (muscle) | Kerbe (notch) | Hafen (port) |
| Banane (banana) | Aprikose (apricot) | Streichholz (match) | Banjo (banjo) | Schwager (brother-in-law) |
| Diamant (diamond) | Rubin (ruby) | Mücke (mosquito) | Dialekt (dialect) | Kruste (crust) |
| Krücke (crutch) | Stock (stick) | Ameise (ant) | Kruste (crust) | Panther (panther) |
| Mond (moon) | Sonne (sun) | Huhn (chicken) | Monitor (monitor) | Biene (bee) |
| Koffer (suitcase) | Tasche (bag) | Stift (pen) | Koffein (caffeine) | Adel (nobility) |
| Hirsch (deer) | Elch (moose) | Säge (saw) | Hirn (brain) | Akt (act) |
| Kiwi (kiwi) | Zitrone (lemon) | Tasche (bag) | Kino (movie theater) | Armee (army) |
| Ratte (rat) | Maus (mouse) | Füller (pen) | Raster (grid) | Pickel (pimple) |
| Kühlschrank (refrigerator) | Herd (stove) | Saxophon (saxophone) | Kübel (bucket) | Zwiesel (zwiesel) |
| Geist (ghost) | Zombie (zombie) | Floß (raft) | Geier (vulture) | Konsum (consumption) |
| Olive (olive) | Bohne (bean) | Lineal (ruler) | Olympia (olympics) | Zirkus (circus) |
| Domino (domino) | Würfel (dice) | Papaya (papaya) | Dominanz (dominance) | Zeichen (sign) |
| Kokosnuss (coconut) | Papaya (papaya) | Deodorant (deodorant) | Kokain (cocaine) | Analyse (analysis) |
| Regen (rain) | Schnee (snow) | Mütze (cap) | Regel (rule) | Mikado (mikado) |
| Zirkel (divider) | Lineal (ruler) | Flasche (bottle) | Zirkus (circus) | Seide (silk) |
| Rock (skirt) | Hemd (shirt) | Gras (grass) | Rost (rust) | Hirn (brain) |
| Trompete (trumpet) | Saxophon (saxophone) | Schnee (snow) | Troll (troll) | Kokain (cocaine) |
| Kappe (cap) | Mütze (cap) | Würfel (dice) | Kapitän (captain) | Troll (troll) |
| Drucker (printer) | Scanner (scanner) | Zombie (zombie) | Druide (druid) | Hotel (hotel) |
| Bumerang (boomerang) | Frisbee (frisbee) | Aprikose (apricot) | Bulle (bull) | Druide (druid) |
| Hammer (hammer) | Zange (Pliers) | Traube (grape) | Hamster (hamster) | Bulle (bull) |
| Schnuller (pacifier) | Flasche (bottle) | Sessel (armchair) | Schnur (line) | Marine (navy) |
| Pinsel (brush) | Füller (pen) | Jacuzzi (jacuzzi) | Pickel (pimple) | Bote (messenger) |
| Bier (beer) | Wein (wine) | Kamm (comb) | Biene (bee) | Tante (aunt) |
| Salzstreuer (salt shaker) | Pfeffermühle (pepper grinder) | Pistole (pistol) | Salbe (ointment) | Pudel (poodle) |
| Schaufel (shovel) | Hacke (pickaxe) | Ziege (goat) | Schau (show) | Kerbe (notch) |
| Baguette (baguette) | Schrippe (bread roll) | Fuß (foot) | Bagger (excavator) | Hain (grove) |
| Hase (rabbit) | Wiesel (weasel) | Schilf (reed) | Hafen (port) | Rost (rust) |
| Bett (bed) | Sofa (sofa) | Granate (grenade) | Berg (mountain) | Tuba (tuba) |
| Chamäleon (chameleon) | Eidechse (lizard) | Pfeffermühle (pepper grinder) | Chaos (chaos) | Olympia (olympics) |
| Karotte (carrot) | Gurke (cucumber) | Eidechse (lizard) | Kardinal (cardinal) | Monitor (monitor) |
| Tunnel (tunnel) | Höhle (cave) | Harfe (harp) | Tuba (tuba) | Kanister (canister) |
| Kompass (compass) | Uhr (clock) | Melone (melon) | Konsum (consumption) | Beere (berry) |
| Dinosaurier (dinosaurs) | Mammut (mammoth) | Badewanne (bathtub) | Dimension (dimension) | Mediation (mediation) |
| Strauß (ostrich) | Huhn (chicken) | Knie (knee) | Strauch (shrub) | Kübel (bucket) |
| Faden (thread) | Wolle (wool) | Gurke (cucumber) | Fahne (banner) | Kirche (church) |
| Blume (flower) | Gras (grass) | Trophäe (trophy) | Bluse (blouse) | Pixel (pixel) |
| Parfüm (perfume) | Deodorant (deodorant) | Höhle (cave) | Park (park) | Brigade (brigade) |
| Kuh (cow) | Ziege (goat) | Dose (can) | Kuli (pen) | Fahne (banner) |
| Geschenk (gift) | Paket (package) | Sofa (sofa) | Geschichte (story) | Spitzel (spy) |
| Schmetterling (butterfly) | Libelle (dragonfly) | Marmelade (jam) | Schmerz (pain) | Dominanz (dominance) |
| Gitarre (guitar) | Harfe (harp) | Libelle (dragonfly) | Giraffe (giraffe) | Kreis (circle) |
| Tanne (fir) | Fichte (spruce) | Messer (knife) | Tante (aunt) | Regel (rule) |
| Pool (pool) | Jacuzzi (jacuzzi) | Gans (goose) | Pudel (poodle) | Ehre (honor) |
| Bombe (bomb) | Granate (grenade) | Wiesel (weasel) | Borste (bristle) | Filz (felt) |
| Antenne (antenna) | Radar (radar) | Jalousie (louvre) | Antike (antiquity) | Kardinal (cardinal) |
| Finger (finger) | Zeh (toe) | Puppe (doll) | Filz (felt) | Elfe (elf) |
| Vorhang (curtain) | Jalousie (louvre) | Mammut (mammoth) | Vorort (suburb) | Kredit (credit) |
| Thron (throne) | Sessel (armchair) | Hemd (shirt) | Trost (consolation) | Raster (grid) |
| Axt (axe) | Säge (saw) | Jeep (jeep) | Akt (act) | Lehm (clay) |
| Krebs (crab) | Hummer (lobster) | Uhr (clock) | Kredit (credit) | Schau (show) |
| Kanone (cannon) | Pistole (pistol) | Tablette (tablet) | Kanister (canister) | Vorort (suburb) |
| Ananas (pineapple) | Melone (melon) | Fernglas (binoculars) | Analyse (analysis) | Bluse (blouse) |
| Fliege (fly) | Mücke (mosquito) | Scanner (scanner) | Fliese (tile) | Dialekt (dialect) |
| Arm (arm) | Fuß (foot) | Wein (wine) | Armee (army) | Fliese (tile) |
| Rose (rose) | Tulpe (tulip) | Falke (falcon) | Rosine (raisin) | Fund (discovery) |
| Boot (boat) | Floß (raft) | Maus (mouse) | Bote (messenger) | Schnur (line) |
| Pizza (pizza) | Kuchen (cake) | Hacke (pickaxe) | Pixel (pixel) | Hantel (dumbbell) |
| Tasse (cup) | Dose (can) | Elch (moose) | Taste (button) | Schmerz (pain) |
| Fuchs (fox) | Wolf (wolf) | Zeh (toe) | Fund (discovery) | Gabe (gift) |
| Schwan (swan) | Gans (goose) | Zitrone (lemon) | Schwager (brother-in-law) | Borste (bristle) |
| Hai (shark) | Wal (whale) | Buch (book) | Hain (grove) | Berg (mountain) |
| Marionette (puppet) | Puppe (doll) | Knoblauch (garlic) | Marine (navy) | Koffein (caffeine) |
| Pille (pill) | Tablette (tablet) | Wolf (wolf) | Pilot (pilot) | Strauch (shrub) |
| Feuerzeug (lighter) | Streichholz (match) | Hummer (lobster) | Feudalismus (feudalism) | Kapitän (captain) |
| Handschuh (glove) | Strumpf (stocking) | Fichte (spruce) | Hantel (dumbbell) | Rosine (raisin) |
| Besen (broom) | Harke (rake) | Mandel (almond) | Beere (berry) | Geier (vulture) |
| Honig (honey) | Marmelade (jam) | Speer (spear) | Hotel (hotel) | Park (park) |
| Seife (soap) | Duschgel (shower gel) | Fackel (torch) | Seide (silk) | Klerus (clergy) |
| Panzer (tank) | Jeep (jeep) | Sonne (sun) | Panther (panther) | Boden (ground) |
| Kreide (chalk) | Stift (pen) | Tulpe (tulip) | Kreis (circle) | Mantra (mantra) |
| Mantel (coat) | Anorak (jacket) | Bohne (bean) | Mantra (mantra) | Bürde (burden) |
| Medaille (medal) | Trophäe (trophy) | Anorak (jacket) | Mediation (mediation) | Kuli (pen) |
| Gabel (fork) | Messer (knife) | Tiger (tiger) | Gabe (gift) | Knospe (bud) |
| Bogen (bow) | Speer (spear) | Wolle (wool) | Boden (ground) | Giraffe (giraffe) |
| Erdnuß (peanut) | Mandel (almond) | Harke (rake) | Ehre (honor) | Pilot (pilot) |
| Kirsche (cherry) | Traube (grape) | Paket (package) | Kirche (church) | Bagger (excavator) |
| Adler (eagle) | Falke (falcon) | Schrippe (bread roll) | Adel (nobility) | Kino (movie theater) |
| Aquarium (aquarium) | Käfig (cage) | Reis (rice) | Aquarell (watercolor) | Liste (list) |
| Drachen (kite) | Ballon (balloon) | Kabel (electric wire) | Draht (wire) | Kehle (throat) |

| Kartoffel (potato) | Reis (rice) | Augenbinde (blindfold) | Karton (carton) | Schlaufe (loop) |
| --- | --- | --- | --- | --- |
| Kegel (pin) | Dart (darts) | Auge (eye) | Kehle (throat) | Mast (mast) |
| Lippen (lips) | Auge (eye) | Käfig (cage) | Liste (list) | Draht (wire) |
| Maske (mask) | Augenbinde (blindfold) | Kralle (claw) | Mast (mast) | Schnaps (schnapps) |
| Schlauch (hose) | Kabel (electric wire) | Dart (darts) | Schlaufe (loop) | Karton (carton) |
| Schnabel (beak) | Kralle (claw) | Ballon (balloon) | Schnaps (schnapps) | Aquarell (watercolor) |

*Appendix 2. Distribution of response latencies in the different conditions and tasks*

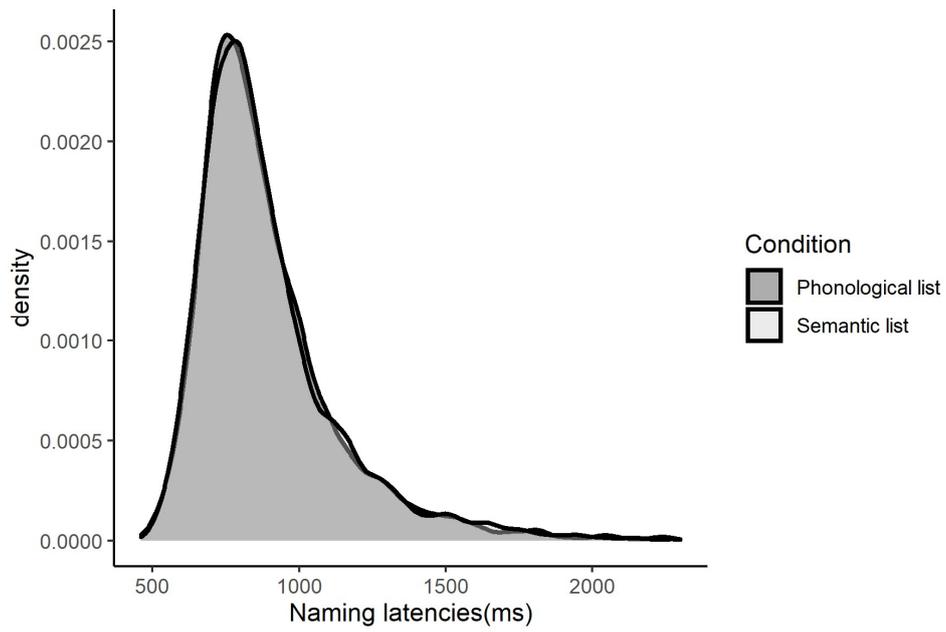

Figure.A.1. Distribution of naming latencies in the unrelated conditions used in the phonological and semantic contrasts.

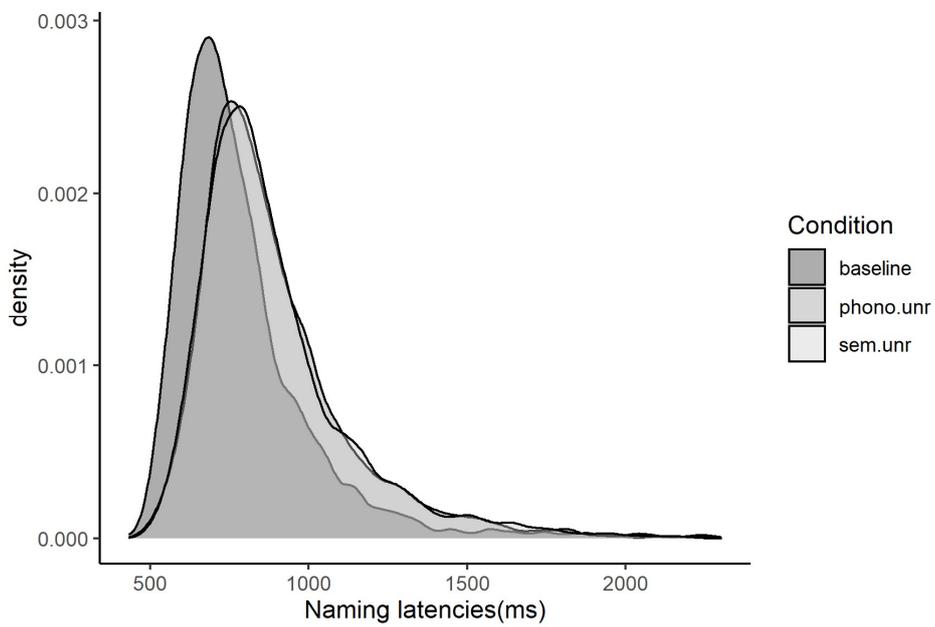

Figure.A.2. Distribution of naming latencies in the baseline and two unrelated conditions

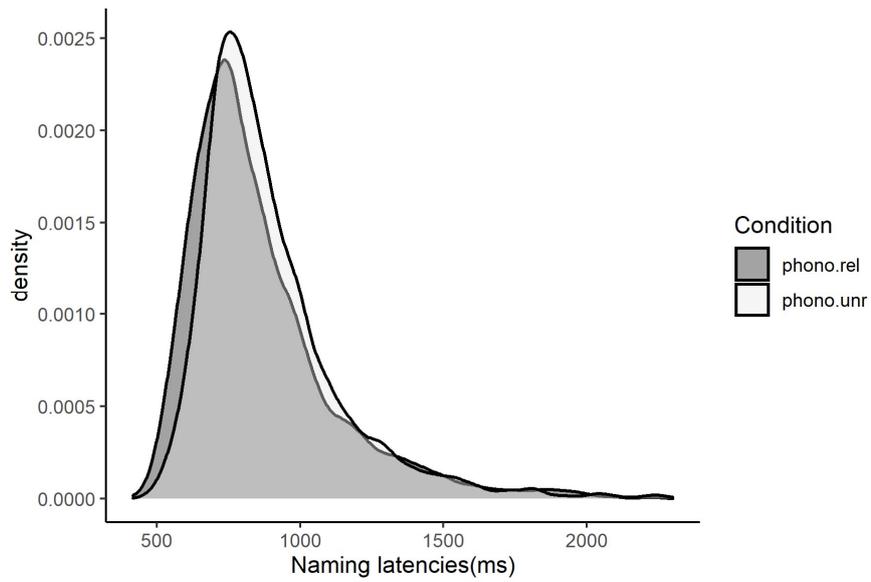

Figure.A.3. Distribution of naming latencies in the phonologically related and unrelated conditions

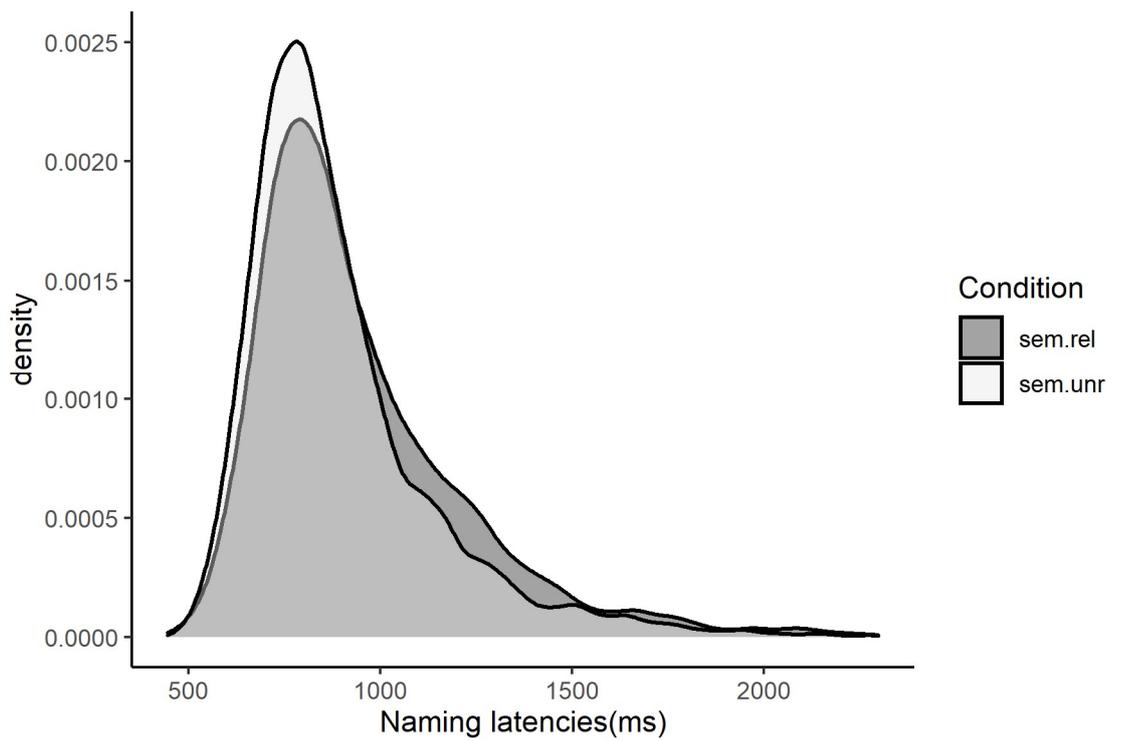

Figure.A.4. Distribution of naming latencies in the semantically related and unrelated conditions

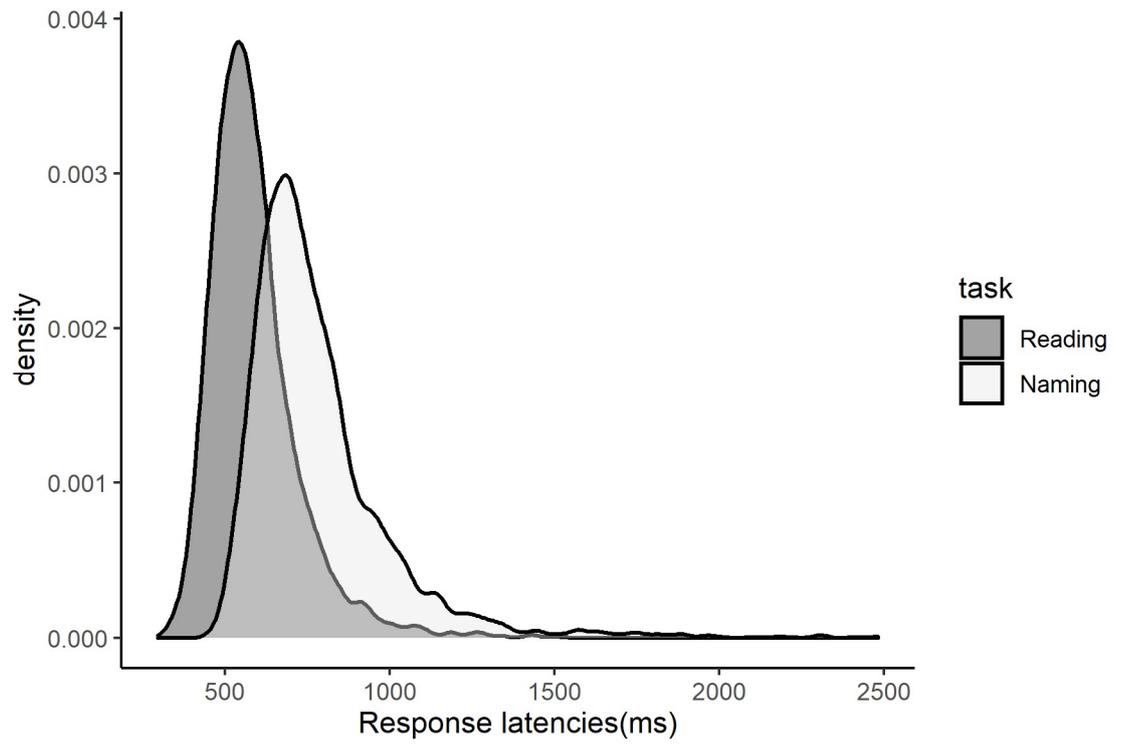

Figure.A.5. Distribution of naming latencies in the baseline condition and in the reading aloud task

*Appendix 3. Output of linear mixed-effects model with the four experimental contrasts and their interactions with trial position and number of repetitions (additional analysis)*

|  | RT | | |
|---|---|---|---|
|  | *Estimates* | *CI* | *p* |
| (Intercept) | 773.83 | 743.89 , 803.77 | <0.001 |
| General interference contrast 1 | 97.74 | 89.68 , 105.79 | <0.001 |
| General interference contrast 2 | 98.08 | 88.47 , 107.68 | <0.001 |
| Phonological contrast | -33.36 | -47.63 , -19.09 | <0.001 |
| Semantic contrast | 54.56 | 38.80 , 70.31 | <0.001 |
| Number of Repetitions (centered) | -44.40 | -62.31 , -26.49 | <0.001 |
| Position of trial in experiment (centered, linear) | 4020.99 | 503.99 , 7537.99 | 0.025 |
| Position of trial in experiment (centered, quadratic) | 1878.83 | 1009.31 , 2748.35 | <0.001 |
| Phonological contrast * Number of Repetitions (centered) | -3.67 | -29.21 , 21.88 | 0.779 |
| General interference contrast 1 * Number of Repetitions (centered) | -13.51 | -39.01 , 11.99 | 0.299 |
| General interference contrast 2 * Number of Repetitions (centered) | -9.68 | -35.21 , 15.85 | 0.457 |
| Semantic contrast * Number of Repetitions (centered) | -24.62 | -50.50 , 1.26 | 0.062 |
| Semantic contrast * Position of trial in experiment (centered, linear) | 3506.92 | -1450.60 , 8464.45 | 0.166 |
| Semantic contrast * Position of trial in experiment (centered, quadratic) | 1376.78 | 390.93 , 2362.63 | 0.006 |
| Phonological contrast * Position of trial in experiment (centered, linear) | 1900.22 | -3015.07 , 6815.51 | 0.449 |
| Phonological contrast * Position of trial in experiment (centered, quadratic) | 85.20 | -904.02 , 1074.41 | 0.866 |
| General interference contrast 1* Position of trial in experiment (centered, linear) | 2252.54 | -2635.26 , 7140.34 | 0.366 |
| General interference contrast 1* Position of trial in experiment (centered, quadratic) | 46.54 | -927.81 , 1020.89 | 0.925 |
| General interference contrast 1* Position of trial in experiment (centered, linear) | 1588.19 | -3288.84 , 6465.22 | 0.523 |

| General interference contrast 2 * Position of trial in experiment (centered, quadratic) | -268.72 | -1238.21 , 700.76 | 0.587 |

*Appendix 4. Output of four linear mixed-effects models testing the interactions between each of the experimental contrast and by participant quintiles (Analysis 3).*

| Predictor | Estimate | 95% Conf. Int. | p |
| --- | --- | --- | --- |
| | Model 1. Phonological facilitation * Quintile | | |
| (Intercept) | 770.53 | 742.60 , 798.46 | <0.001 |
| Number of Repetitions (centered) | 0.41 | -0.40 , 1.23 | 0.322 |
| Phonological contrast | -35.01 | -45.65 , -24.37 | <0.001 |
| Quintile (linear) | 21187.57 | 21007.48 , 21367.66 | <0.001 |
| Quintile (quadratic) | 4924.53 | 4752.80 , 5096.27 | <0.001 |
| Quintile (cubic) | 2619.52 | 2448.99 , 2790.04 | <0.001 |
| Semantic contrast | 52.98 | 40.99 , 64.96 | <0.001 |
| General contrast 1 | 100.68 | 90.46 , 110.89 | <0.001 |
| General contrast 2 | 100.62 | 90.64 , 110.59 | <0.001 |
| Phonological facilitation * Quintile (linear) | 1474.10 | 1092.38 , 1855.82 | <0.001 |
| Phonological facilitation * Quintile (quadratic) | 196.18 | -184.62 , 576.99 | 0.313 |
| Phonological facilitation * Quintile (cubic) | 336.19 | -44.00 , 716.38 | 0.083 |

|  | Model 2. Semantic interference * Quintile | | |
| --- | --- | --- | --- |
| (Intercept) | 776.51 | 748.07 , 804.94 | <0.001 |
| Number of Repetitions (centered) | -1.00 | -1.66 , -0.34 | 0.003 |
| Phonological contrast | -35.64 | -46.15 , -25.14 | <0.001 |
| Quintile (linear) | 20830.72 | 19268.85 , 22392.58 | <0.001 |
| Quintile (quadratic) | 5198.70 | 4593.41 , 5803.99 | <0.001 |
| Quintile (cubic) | 2804.33 | 2517.81 , 3090.86 | <0.001 |
| Semantic contrast | 56.27 | 44.55 , 67.99 | <0.001 |
| General contrast 1 | 99.62 | 89.19 , 110.04 | <0.001 |
| General contrast 2 | 99.39 | 89.32 , 109.47 | <0.001 |
| Semantic contrast * Quintile (linear) | 4257.87 | 3940.99 , 4574.75 | <0.001 |
| Semantic contrast * Quintile (quadratic) | 421.06 | 109.33 , 732.79 | 0.008 |
| Semantic contrast * Quintile (cubic) | -136.43 | -444.90 , 172.04 | 0.386 |
|  | Model 3. General interference 1 * Quintile | | |
| (Intercept) | 777.44 | 748.98 , 805.89 | <0.001 |
| Number of Repetitions (centered) | -1.18 | -1.85 , -0.50 | 0.001 |
| Phonological contrast | -36.19 | -46.53 , -25.84 | <0.001 |
| Quintile (linear) | 21345.83 | 19701.13 , 22990.52 | <0.001 |
| Quintile (quadratic) | 5191.49 | 4551.53 , 5831.45 | <0.001 |
| Quintile (cubic) | 2648.07 | 2344.37 , 2951.77 | <0.001 |
| Semantic contrast | 53.17 | 41.10 , 65.25 | <0.001 |
| General interference contrast 1 | 100.47 | 90.41 , 110.54 | <0.001 |
| General interference contrast 2 | 99.03 | 89.10 , 108.95 | <0.001 |
| General interference contrast 1 * Quintile (linear) | 696.52 | 437.42 , 955.62 | <0.001 |
| General interference contrast 1 * Quintile (quadratic) | 167.41 | -87.21 , 422.03 | 0.198 |
| General interference contrast 1 * Quintile (cubic) | 287.09 | 34.19 , 539.99 | 0.026 |
| | Model 4. General interference 2 * Quintile | | |
| (Intercept) | 776.62 | 748.35 , 804.89 | <0.001 |
| Number of Repetitions (centered) | -1.17 | -1.84 , -0.50 | 0.001 |
| Phonological contrast | -35.26 | -45.63 , -24.90 | <0.001 |
| Quintile (linear) | 20490.79 | 18808.41 , 22173.17 | <0.001 |
| Quintile (quadratic) | 5008.98 | 4356.32 , 5661.63 | <0.001 |
| Quintile (cubic) | 2725.78 | 2414.88 , 3036.67 | <0.001 |
| Semantic contrast | 52.77 | 40.86 , 64.68 | <0.001 |
| General interference contrast 1 | 99.34 | 88.92 , 109.76 | <0.001 |
| General interference contrast 2 | 102.33 | 92.21 , 112.46 | <0.001 |
| General interference contrast 2 * Quintile (linear) | 2926.47 | 2668.74 , 3184.19 | <0.001 |
| General interference contrast 2 * Quintile (quadratic) | 736.63 | 483.40 , 989.85 | <0.001 |
| General interference contrast 2 * Quintile (cubic) | 156.90 | -93.97 , 407.76 | 0.220 |

*Appendix 5. Naming latencies, as predicted by the statistical models of analysis 3, for each experimental contrast*

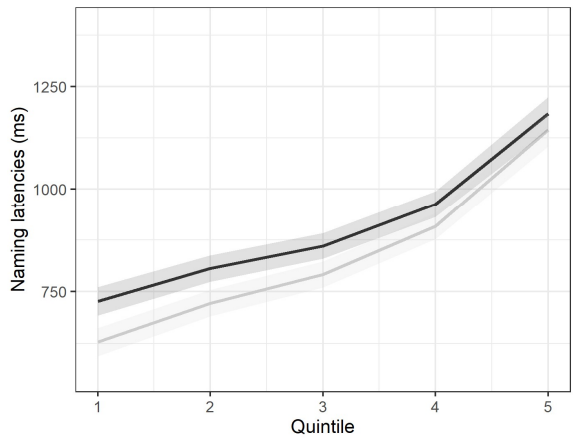

Phonological facilitation

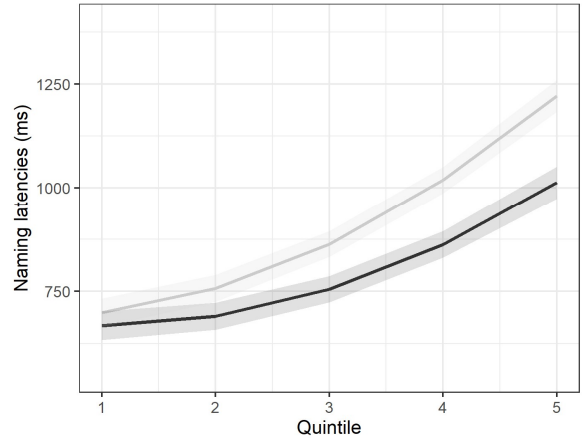

Semantic interference

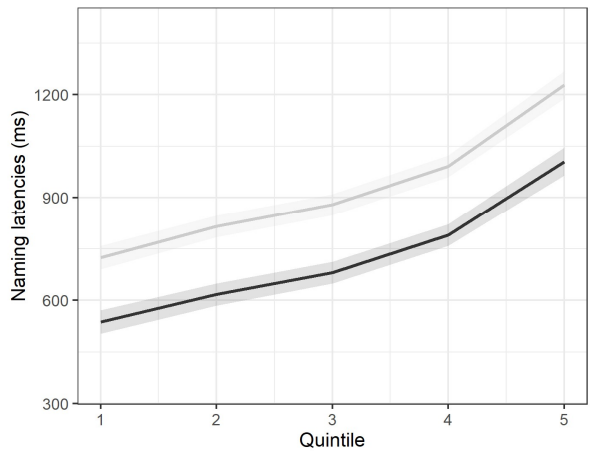

General interference 1

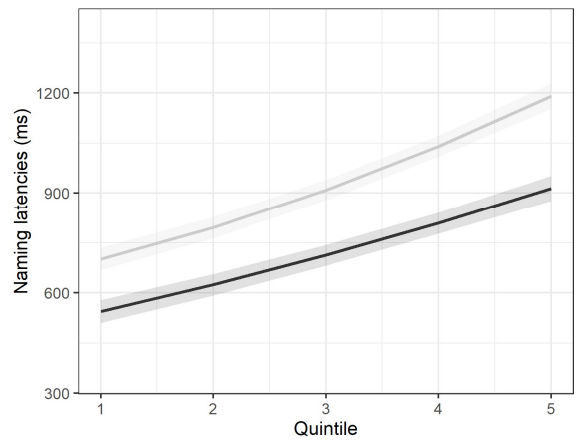

General interference 2

*Appendix 6. Output of linear mixed-effects models for each by participant quintile (Analysis 3)*

|  | ***Estimate*** | ***95% Conf. Int.*** | ***p*** |
|---|---|---|---|
| Model 1. Quintile 1 ||||
| (Intercept) | 628.37 | 610.12 , 646.62 | <0.001 |
| Number of Repetitions (centered) | -2.45 | -3.37 , -1.52 | <0.001 |
| Phonological contrast | -50.67 | -57.62 , -43.72 | <0.001 |
| Semantic contrast | 22.27 | 15.80 , 28.74 | <0.001 |
| General interference contrast 1 | 73.68 | 66.22 , 81.15 | <0.001 |
| General interference contrast 2 | 72.20 | 64.78 , 79.61 | <0.001 |
| Model 2. Quintile 2 ||||
| (Intercept) | 687.61 | 665.61 , 709.61 | <0.001 |
| Number of Repetitions (centered) | -0.94 | -1.45 , -0.42 | <0.001 |
| Phonological contrast | -42.01 | -52.00 , -32.01 | <0.001 |
| Semantic contrast | 35.41 | 26.71 , 44.12 | <0.001 |
| General interference contrast 1 | 87.21 | 77.50 , 96.92 | <0.001 |
| General interference contrast 2 | 83.99 | 74.47 , 93.51 | <0.001 |
| Model 3. Quintile 3 ||||
| (Intercept) | 745.98 | 720.26 , 771.71 | <0.001 |
| Number of Repetitions (centered) | -0.36 | -0.94 , 0.21 | 0.214 |
| Phonological contrast | -34.16 | -46.49 , -21.84 | <0.001 |
| Semantic contrast | 57.94 | 41.04 , 74.84 | <0.001 |
| General interference contrast 1 | 95.88 | 84.83 , 106.93 | <0.001 |
| General interference contrast 2 | 93.59 | 82.31 , 104.88 | <0.001 |

| Model 4. Quintile 4 | | | |
|---|---|---|---|
| (Intercept) | 829.36 | 797.56 , 861.15 | <0.001 |
| Number of Repetitions (centered) | -1.53 | -2.50 , -0.56 | 0.002 |
| Phonological contrast | -31.67 | -47.16 , -16.19 | <0.001 |
| Semantic contrast | 77.95 | 58.54 , 97.37 | <0.001 |
| General interference contrast 1 | 116.56 | 100.99 , 132.14 | <0.001 |
| General interference contrast 2 | 114.24 | 99.68 , 128.81 | <0.001 |

| Model 5. Quintile 5 | | | |
|---|---|---|---|
| (Intercept) | 1058.90 | 1004.29 , 1113.51 | <0.001 |
| Number of Repetitions (centered) | -0.78 | -4.91 , 3.35 | 0.710 |
| Phonological contrast | -11.06 | -37.67 , 15.55 | 0.419 |
| Semantic contrast | 93.93 | 65.97 , 121.88 | <0.001 |
| General interference contrast 1 | 150.37 | 128.64 , 172.11 | <0.001 |
| General interference contrast 2 | 159.71 | 141.03 , 178.39 | <0.001 |

*Appendix 7. Output of four linear mixed-effects models testing the interactions between each of the experimental contrast and by item quintiles (Analysis 3).*

|  | *Estimate* | *95% Conf. Int.* | *p* |
| --- | --- | --- | --- |
|  | Model 1. Phonological contrast * Quintile | | |
| (Intercept) | 774.83 | 760.71 , 788.94 | <0.001 |
| Number of Repetitions (centered) | -1.24 | -1.99 , -0.48 | 0.001 |
| Phonological contrast | -34.83 | -48.76 , -20.90 | <0.001 |
| Quintile (linear) | 21907.54 | 21065.68 , 22749.40 | <0.001 |
| Quintile (quadratic) | 5226.22 | 4653.96 , 5798.49 | <0.001 |
| Quintile (cubic) | 2444.48 | 2196.46 , 2692.50 | <0.001 |
| Semantic contrast | 51.95 | 38.23 , 65.66 | <0.001 |
| General interference contrast 1 | 103.30 | 93.70 , 112.90 | <0.001 |
| General interference contrast 2 | 102.01 | 91.57 , 112.44 | <0.001 |
| Phonological contrast * Quintile (linear) | 1590.71 | 1231.93 , 1949.48 | <0.001 |
| Phonological contrast * Quintile (quadratic) | 399.82 | 50.52 , 749.12 | 0.025 |
| Phonological contrast * Quintile (cubic) | 302.50 | -43.79 , 648.80 | 0.087 |
|  | Model 2. Semantic contrast * Quintile | | |
| (Intercept) | 773.91 | 759.79 , 788.04 | <0.001 |
| Number of Repetitions (centered) | -1.24 | -2.04 , -0.43 | 0.003 |
| Phonological contrast | -35.99 | -49.87 , -22.11 | <0.001 |
| Quintile (linear) | 21380.28 | 21180.42 , 21580.14 | <0.001 |
| Quintile (quadratic) | 5194.20 | 4660.03 , 5728.37 | <0.001 |
| Quintile (cubic) | 2596.28 | 2357.65 , 2834.92 | <0.001 |
| Semantic contrast | 54.17 | 41.39 , 66.95 | <0.001 |
| General interference contrast 1 | 103.53 | 93.92 , 113.15 | <0.001 |
| General interference contrast 2 | 102.23 | 91.99 , 112.47 | <0.001 |
| Semantic contrast * Quintile (linear) | 4363.51 | 3977.92 , 4749.10 | <0.001 |
| Semantic contrast * Quintile (quadratic) | 416.13 | 38.79 , 793.47 | 0.031 |
| Semantic contrast * Quintile (cubic) | -232.44 | -606.75 , 141.88 | 0.224 |
| | Model 3. General interference 1 * Quintile | | |
| (Intercept) | 774.76 | 760.79 , 788.72 | <0.001 |
| Number of Repetitions (centered) | -1.41 | -2.16 , -0.66 | <0.001 |
| Phonological contrast | -35.75 | -49.66 , -21.84 | <0.001 |
| Quintile (linear) | 21887.18 | 20920.53 , 22853.84 | <0.001 |
| Quintile (quadratic) | 5169.55 | 4646.64 , 5692.46 | <0.001 |
| Quintile (cubic) | 2445.20 | 2118.72 , 2771.68 | <0.001 |
| Semantic contrast | 51.76 | 38.03 , 65.50 | <0.001 |
| General interference contrast 1 | 103.74 | 94.23 , 113.25 | <0.001 |
| General interference contrast 2 | 102.01 | 91.58 , 112.44 | <0.001 |
| General interference contrast 1 * Quintile (linear) | 930.85 | 622.76 , 1238.93 | <0.001 |
| General interference contrast 1 * Quintile (quadratic) | 199.70 | -87.62 , 487.02 | 0.173 |
| General interference contrast 1 * Quintile (cubic) | 120.47 | -160.88 , 401.81 | 0.401 |

| Model 4. General interference 2 * Quintile | | | |
|---|---|---|---|
| (Intercept) | 773.64 | 759.73 , 787.54 | <0.001 |
| Number of Repetitions (centered) | -1.28 | -2.03 , -0.54 | 0.001 |
| Phonological contrast | -35.11 | -49.14 , -21.07 | <0.001 |
| Quintile (linear) | 21190.66 | 20225.58 , 22155.75 | <0.001 |
| Quintile (quadratic) | 5078.18 | 4606.68 , 5549.67 | <0.001 |
| Quintile (cubic) | 2435.77 | 2136.70 , 2734.84 | <0.001 |
| Semantic contrast | 51.61 | 38.18 , 65.03 | <0.001 |
| General interference contrast 1 | 103.10 | 93.20 , 112.99 | <0.001 |
| General interference contrast 2 | 103.83 | 93.79 , 113.86 | <0.001 |
| General interference contrast 2 * Quintile (linear) | 2788.54 | 2490.11 , 3086.97 | <0.001 |
| General interference contrast 2 * Quintile (quadratic) | 455.68 | 171.85 , 739.51 | 0.002 |
| General interference contrast 2 * Quintile (cubic) | 182.83 | -96.56 , 462.23 | 0.200 |

*Appendix 8. Output of linear mixed-effects models for each by item quintile (Analysis 3).*

|  | *Estimate* | *95% Conf. Int.* | *p* |
|---|---|---|---|
| | Model 1. Quintile 1 | | |
| (Intercept) | 630.43 | 619.40 , 641.47 | <0.001 |
| Number of Repetitions (centered) | -5.18 | -6.15 , -4.21 | <0.001 |
| Phonological contrast | -45.76 | -54.07 , -37.46 | <0.001 |
| Semantic contrast | 18.67 | 11.32 , 26.02 | <0.001 |
| General interference contrast 1 | 72.13 | 66.20 , 78.07 | <0.001 |
| General interference contrast 2 | 74.57 | 67.84 , 81.30 | <0.001 |
| | Model 2. Quintile 2 | | |
| (Intercept) | 681.55 | 670.49 , 692.61 | <0.001 |
| Number of Repetitions (centered) | -1.25 | -1.85 , -0.64 | <0.001 |
| Phonological contrast | -40.27 | -51.74 , -28.80 | <0.001 |
| Semantic contrast | 28.82 | 18.47 , 39.17 | <0.001 |
| General interference contrast 1 | 85.70 | 78.17 , 93.23 | <0.001 |
| General interference contrast 2 | 86.90 | 78.18 , 95.62 | <0.001 |
| | Model 3. Quintile 3 | | |
| (Intercept) | 742.94 | 729.37 , 756.52 | <0.001 |
| Number of Repetitions (centered) | -1.04 | -1.71 , -0.38 | 0.002 |
| Phonological contrast | -38.46 | -53.67 , -23.24 | <0.001 |
| Semantic contrast | 52.83 | 37.37 , 68.29 | <0.001 |
| General interference contrast 1 | 101.97 | 92.28 , 111.67 | <0.001 |
| General interference contrast 2 | 98.23 | 87.77 , 108.69 | <0.001 |
| Model 4. Quintile 4 | | | |
| (Intercept) | 832.83 | 815.58 , 850.09 | <0.001 |
| Number of Repetitions (centered) | -1.85 | -2.93 , -0.77 | 0.001 |
| Phonological contrast | -31.25 | -51.15 , -11.35 | 0.003 |
| Semantic contrast | 81.68 | 61.59 , 101.78 | <0.001 |
| General interference contrast 1 | 120.07 | 106.67 , 133.48 | <0.001 |
| General interference contrast 2 | 111.85 | 97.47 , 126.23 | <0.001 |
| Model 5. Quintile 5 | | | |
| (Intercept) | 1042.95 | 1012.27 , 1073.64 | <0.001 |
| Number of Repetitions (centered) | 4.86 | 0.52 , 9.20 | 0.028 |
| Phonological contrast | -6.41 | -37.99 , 25.18 | 0.692 |
| Semantic contrast | 97.30 | 66.70 , 127.90 | <0.001 |
| General interference contrast 1 | 153.44 | 128.13 , 178.75 | <0.001 |
| General interference contrast 2 | 163.80 | 137.72 , 189.89 | <0.001 |

*Appendix 9. Results of analysis with difference measure (Analysis 2) with data collected with the same design but a different pool of participants*

|  | | RT | |
|---|---|---|---|
| *Predictors* | *Estimates* | *95% Conf. Int.* | *p* |
| (Intercept) | 742.72 | 719.86 , 765.58 | <0.001 |
| Phonological contrast | -44.91 | -67.85 , -21.98 | <0.001 |
| Difference measure (centered, linear) | 597.52 | 235.94 , 959.10 | 0.001 |
| Difference measure (centered, quadratic) | 332.97 | 60.15 , 605.79 | 0.017 |
| Semantic contrast | 27.95 | 16.77 , 39.12 | <0.001 |
| Difference measure (centered, linear) * Semantic contrast | -98.39 | -639.89 , 443.11 | 0.722 |
| Difference measure (centered, quadratic) * Semantic contrast | -678.98 | -1210.74 , -147.22 | 0.012 |
| Difference measure (centered, linear) * Phonological contrast | -127.32 | -693.83 , 439.20 | 0.660 |
| Difference measure (centered, quadratic) * Phonological contrast | -629.70 | -1214.52 , -44.88 | 0.035 |

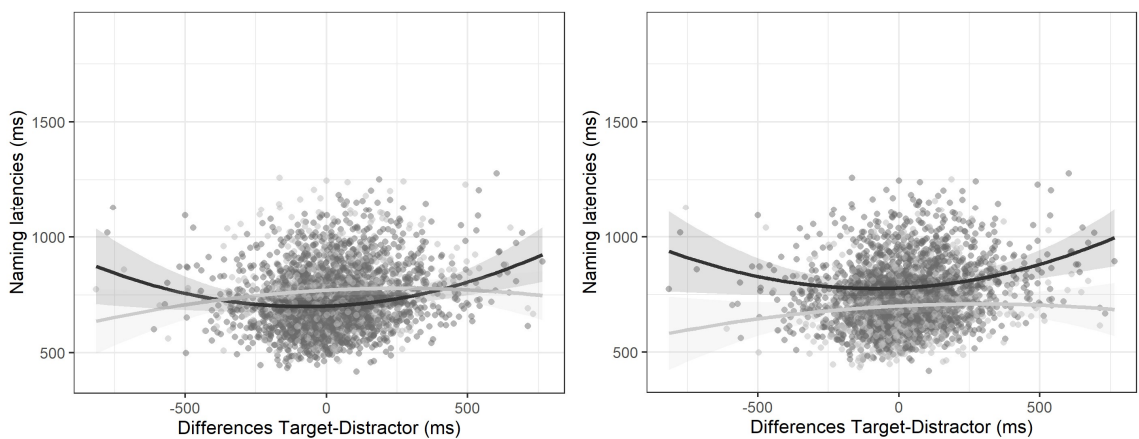

Figure A11. Interactions between the difference measure and the semantic contrast (left) and between the difference measure and the phonological contrast (right) as predicted by the statistical model. Black line = unrelated trials, Grey line = related trials.

*Appendix 10. Priors for Bayes Factors*

| Parameter | Prior 1 | Prior 2 |
|---|---|---|
| Intercept | $N(700, 300)$ | $N(700, 300)$ |
| Phonological contrast | $N(0, 200)$ | $N(0, 200)$ |
| Semantic contrast | $N(0, 200)$ | $N(0, 200)$ |
| General interference contrasts | $N(0, 200)$ | $N(0, 200)$ |
| Trial position (centered) | $N(0, 10)$ | $N(0, 10)$ |
| Repetition (centered) | $N(0, 200)$ | $N(0, 200)$ |
| EEG alpha power | $N(0, 10)$ | $N(0, 1)$ |
| Interaction experimental contrasts * EEG alpha power | $N(0, 10)$ | $N(0, 1)$ |
| Standard deviations in model | $N_+(0, 100)$ | $N_+(0, 100)$ |

Table A10.2. Results of Bayes factor analyses

| Fixed-effect | Prior 1 | Prior 2 |
|---|---|---|
| EEG alpha power | | |
| Phonological contrast * EEG alpha power | | |
| Semantic contrast * EEG alpha power | | |
| General interference contrast 1 * EEG alpha power | | |
| General interference contrast 2 * EEG alpha power | | |

*Appendix 10. Priors for Bayes Factors*